\documentclass[a4paper,12pt]{article}
\usepackage{amsbsy}
\usepackage[fleqn]{amsmath}
\usepackage{amssymb}
\usepackage{array}
\usepackage{booktabs}
\usepackage{xcolor}
\usepackage[colorlinks, linkcolor=blue, anchorcolor=blue, citecolor=blue]{hyperref}
\usepackage{mathrsfs}
\usepackage{tabularx}
\usepackage{latexsym}
\usepackage{textcomp}
\usepackage{pstricks}
\usepackage{colortbl}
\usepackage{multirow}
\usepackage{fancyhdr}
\usepackage{bm}
\usepackage{booktabs}
\usepackage{subfig}
\usepackage{caption}
\usepackage{soul}
\usepackage{cleveref}
\usepackage{cite}
\usepackage{lineno}
\usepackage{ulem}
\usepackage{graphicx}
\usepackage{epstopdf}
\usepackage{indentfirst}
\usepackage{setspace}
\usepackage[latin1]{inputenc}
\usepackage{titlesec}
\usepackage{tabularx} 
\usepackage{ragged2e} 
\usepackage{booktabs} 
\usepackage{qcircuit} 
\usepackage{braket}   
\usepackage{svg}      
\usepackage{amsmath}

\usepackage{algorithm}
\usepackage{algpseudocode}
\usepackage{mdframed}
\usepackage{xcolor}

\titleformat*{\section}{\large\bf}
\titleformat*{\subsection}{\normalsize\bf}
\crefname{section}{Section}{Sections}
\crefname{equation}{Eq.}{Eqs.}
\crefname{figure}{Fig.}{Figs.}
\crefname{table}{Table}{Tables}
\crefname{algorithm}{Algorithm}{Algorithms}
\begin{document}
%
%
\begin{center}
\Large\textbf{Quantum computing with error mitigation for data-driven computational homogenization }
\end{center}
\large{
\begin{center}
\setcounter{footnote}{1}
\textbf{Zengtao Kuang}$^{a,}$\footnote{These authors contributed equally to this work.\label{co}},
\setcounter{footnote}{1}
\textbf{Yongchun Xu}$^{a,}$$\textsuperscript{\ref{co}}$,
\textbf{Qun Huang$^{a}$}, \textbf{Jie Yang$^{a}$}, \\\textbf{Chafik El Kihal$^{a}$}, 
\setcounter{footnote}{0}
\textbf{Heng Hu}$^{a,b,}$\footnote{Corresponding author. E-mail address: huheng@whu.edu.cn.}$^{}$
\end{center}
}
\small{
\begin{center}
$^a$School of Civil Engineering, Wuhan University, 8 South Road of East Lake, Wuchang, 430072 Wuhan, PR China\\
$^b$School of Mathematics and Statistics, Ningxia University, 750021 Yinchuan, PR China
\end{center}
}
\begin{flushleft}
\large\textbf{Abstract}
\end{flushleft}

As a crossover frontier of physics and mechanics, quantum computing is showing its great potential in computational mechanics.
However, quantum hardware noise remains a critical barrier to achieving accurate simulation results due to the limitation of the current hardware. 
In this paper, we integrate error-mitigated quantum computing in data-driven computational {homogenization}, where the zero-noise extrapolation (ZNE) technique is employed to improve the reliability of quantum computing. Specifically, ZNE is utilized to mitigate the quantum hardware noise in two quantum algorithms for distance calculation, namely a Swap-based algorithm and an H-based algorithm, thereby improving the overall accuracy of data-driven computational homogenization.
\textcolor{black}{Multiscale simulations of a 2D  composite L-shaped beam and a 3D composite cylindrical shell} are conducted with the quantum computer simulator Qiskit, and the results validate the effectiveness of the proposed method.  We believe this work presents a promising step towards using quantum computing in computational mechanics.

\begin{flushleft}
\justifying\textbf{Keywords:} Quantum computing;  Data-driven computational {homogenization}; Error mitigation; Zero-noise extrapolation; Distance calculation.
\end{flushleft}

%
%
\section{Introduction}

Due to the unique properties of quantum systems like superposition and entanglement, quantum computing enables exponentially reducing the complexity of computation in certain tasks. It has attracted the increasing attention of researchers in the field of computational mechanics. Until now, quantum computing has been applied to solve, e.g., boundary value problems with finite element method \cite{montanaro2016quantum}, Poisson's equation \cite{raisuddin2022feqa}, fluid dynamics \cite{Meng2023}, composite materials \cite{liu2023towards}, optimization problems \cite{kou2024dynamic} and so on.
Recently, Xu et al. \cite{xu2023quantum} introduced a quantum computing enhanced data-driven method, which was validated both in a quantum computer simulator and a real quantum computer.
However, the noise remains one of the limitations of the current quantum hardware, which prevents taking advantage of quantum computing. In this work, we aim to extend the quantum computing enhanced data-driven method in composite structures, and reduce the error from quantum hardware noise so as to improve the accuracy of quantum computing.


{\color{black}
Data-driven computational homogenization \cite{xu2020Data} is a novel multiscale method for composite structures, grounded in the `model-free' data-driven computational mechanics \cite{Ortiz2016}. In this method, simulations are performed separately at the macroscopic and microscopic scales. At the microscopic scale, material properties of the composites are computed by homogenization on the representative volume element (RVE), where equivalent strain-stress data are obtained and stored in a material database. At the macroscopic scale, the data-driven simulation of the macrostructure is carried out directly using the pre-existing material data from the database, eliminating the need for real-time homogenization of the RVE. This approach significantly reduces the computational cost typically associated with linking different scales, compared to the traditional multiscale finite element method (FE$^2$) \cite{raju2021review}. Data-driven computational homogenization has been rapidly adopted in various composite materials and structures, including composite beams \cite{yang2019structural,bai2022data,kuang2023data2,kim2023deep,yang2024data}, composite plates \cite{yan2022data,xu2023material}, and granular materials \cite{karapiperis2021data}, among others.}

%


{\color{black}The computational bottleneck of data-driven simulation at the macroscopic scale arises from the large number of distance calculations. Data-driven simulation relies on iteratively minimizing the distance between material data and conservation laws \cite{Ortiz2016}, which requires nearest-neighbor searches in the material database based on distance calculations.} Specifically, if a nearest-neighbor search is performed in a material database with $N$ data points, each having dimension $D$, then $N$ distance calculations are required, each with a cost of $O(D)$, leading to a total computational complexity of $O(ND)$. When involving a high-dimensional and high-density material database, the distance calculations become a significant computational bottleneck. For example, in a 3D elastic solid problem with $10^6$ material data points, the nearest-neighbor search consumes more than 90\% of the total computation time \cite{kdtree2}.

The emergence of the quantum computer shows promise in accelerating distance calculations in {data-driven computational homogenization}.
{In the work of Xu et al. \cite{xu2023quantum}, a quantum algorithm is introduced to exponentially reduce the complexity of distance calculations from $O(D)$ to $O(\log D)$, therefore improving the overall performance of data-driven computing.
However, they assume the usage of a fault-tolerant quantum computer \cite{shor1996fault,sun2022optical}, and quantum hardware noise is not considered.} {In fact, quantum hardware noise can affect the accuracy of quantum algorithms in distance  calculations, thereby decreasing the overall precision of data-driven computing. This noise issue is particularly evident in data-driven computational homogenization due to the complex constitutive relationships of composite structures, which require more material data and higher accuracy in distance calculations.}
The hardware noise in quantum computers usually originates from unexpected interactions with the environment and from imperfect gate operations \cite{Nielsen}. For instance, environmental disturbances can lead to decoherence in a quantum system, resulting in the loss of information. Meanwhile, unlike classical bits confined to 0 and 1, quantum bits (qubits) in quantum computers can be any superposition of $|0\rangle$ and $|1\rangle$. This makes quantum computers prone to output inaccurate information even with small disturbance \cite{national2019quantum}.
Therefore, addressing the errors caused by the hardware noise is a crucial issue for the application of quantum computing.

The primary objective of this paper is to address the quantum hardware noise issue, thereby improving the reliability of quantum computing for its practical application in data-driven computational homogenization. Currently, there are two kinds of quantum algorithms that can be considered to achieve the goal \cite{national2019quantum}. The first is quantum error correction (QEC), which offers a solution by redundantly encoding quantum information with additional qubits and actively detecting and correcting errors \cite{Nielsen,lidar2013quantum}, resulting in fault-tolerant computing. Unfortunately, the hardware level of quantum computers is now in the era named Noisy Intermediate-Scale Quantum (NISQ) \cite{preskill2018quantum}, meaning the number of qubits is limited. However, QEC requires a large overhead number of additional qubits, making it infeasible for NISQ quantum computers \cite{national2019quantum,preskill2018quantum}. 
The second approach, quantum error mitigation (QEM), emerges as a more practical alternative for NISQ devices \cite{endo2018practical}. 
One prominent QEM method is zero-noise extrapolation (ZNE), which intentionally introduces additional noise and extrapolates results to the zero-noise limit \cite{temme2017error, li2017efficient}. 
An advantage of ZNE is its independence from additional qubits, rendering it well-suited for NISQ quantum computers.
ZNE has been able to improve the accuracy of a noisy superconducting quantum processor to an inaccessible level \cite{kandala2019error}. Notably, IBM Quantum recently demonstrated the utility of NISQ quantum computers before the fault-tolerant era with the help of ZNE, which passes a calculation milestone in the field of quantum computing \cite{kim2023evidence,castelvecchi2023ibm}.

In this paper, {we aim to use quantum computing to reduce the complexity of distance calculation in data-driven computational homogenization, and employ ZNE to mitigate the quantum hardware noise}. 
Two quantum algorithms for distance calculations are presented, i.e., an algorithm based on swap test (Swap-based) \cite{lloyd2013quantum,xu2023quantum} and an algorithm based on one Hadamard gate (H-based) \cite{moradi2022clinical,blank2020quantum}, and ZNE is employed to improve their accuracy considering quantum hardware noise. Furthermore, a $k$-d tree data structure \cite{friedman1977algorithm,kdtree2,kdtree1,bahmani2021kd} is used to reduce the number of distance calculations, resulting in a more favorable computational complexity. The quantum computer simulator Qiskit \cite{Qiskit} which allows one to execute quantum algorithms on classical computers, is utilized to validate the proposed approach. Several numerical experiments are conducted to confirm the effectiveness of ZNE in improving the accuracy of distance calculation. Moreover, we present \textcolor{black}{ multiscale simulations of a 2D  composite L-shaped beam and a 3D composite cylindrical shell} to illustrate the practical utility of the proposed methodology.

The remainder of this article is laid out as follows. In \cref{Methodology}, we introduce the fundamental formulations of data-driven computational {homogenization} and the two quantum algorithms for distance calculation. Subsequently, the basic formulation of ZNE for mitigating quantum hardware noise is demonstrated. In \cref{sec:Validation}, we present numerical examples to evaluate the performance of the two quantum algorithms considering hardware noise, and validate the effectiveness of ZNE in improving the accuracy of distance calculation. Additionally, we use a roof truss as a case study to emphasize the impact of ZNE on improving the accuracy of data-driven computing. \textcolor{black}{Multiscale simulations of a 2D composite L-shaped beam and a 3D composite cylindrical shell are presented  in \cref{sec:Application,sec: 3d shell}, respectively. Conclusions and discussions are provided in \cref{Conclusion}.}

%
%
\section{Methodology}\label{Methodology}

In \cref{DD}, we briefly introduce the formulations of data-driven computational homogenization. This computing scheme involves computationally expensive distance calculations for nearest-neighbor searches. Therefore, the two quantum algorithms for distance calculation are introduced in \cref{dis_quan} to reduce the computational complexity. Finally, the formulation of ZNE for mitigating the influence of hardware noise is introduced in \cref{sec:ZNE}.

\subsection{Data-driven computational homogenization}\label{DD}


{\color{black}In the framework of data-driven computational homogenization, the microscopic and macroscopic problems are addressed separately \cite{xu2020Data}. The microscopic problem is solved offline to construct the material database, denoted as $\bar{\mathscr{D}}$, with further details provided in \ref{sec: micro}. At the macroscopic scale, the mechanical response is computed using the data-driven method, where the material behavior of the composites is directly retrieved from the material database. Below, we provide a brief introduction to the data-driven method applied at the macroscopic scale.}


For the macroscopic problem, a solid structure is considered. Its configuration is discretized with finite elements. 
At each integration point $e$, the data-driven solver seeks to minimize the distance between the corresponding strain-stress data $\bar{\bm{z}}^{*}_e=(\bar{\bm{\varepsilon}}^{*}_{e}, \bar{\bm{\sigma}}^{*}_{e})$ in the material database $\bar{\mathscr{D}}$ and the admissible strain-stress state $\bar{\bm{z}}_e=(\bar{\bm{\varepsilon}}_{e}, \bar{\bm{\sigma}}_{e})$ that satisfies equilibrium and compatibility.
A distance-based functional formulates this constrained minimization problem by the cost function
\begin{equation}\label{Eq01}
	\Pi=\frac{1}{2}\sum_{e=1}^{m}{w_e}\bar{\mathscr{F}}_e(\bar{\bm{z}}_e,\bar{\bm{z}}_e^*)-\bar{\bm{\eta}}^\text{T}\left(\sum_{e=1}^{m}{w}_e\bar{\textbf{B}}_e\bar{\bm{\sigma}}_e-\bar{\bm{f}}\right)
\end{equation}
where the distance is defined as
\begin{equation}\label{Eq02}
	\bar{\mathscr{F}}_e(\bar{\bm{z}}_e,\bar{\bm{z}}_e^*) = (\bar{\bm{\varepsilon}}_e-\bar{\bm{\varepsilon}}_e^*)^{\text{T}}{\bar{\mathbb{C}}}(\bar{\bm{\varepsilon}}_e-\bar{\bm{\varepsilon}}_e^*) + (\bar{\bm{\sigma}}_e-\bar{\bm{\sigma}}_e^*)^{\text{T}}{\bar{\mathbb{C}}}^{-1}(\bar{\bm{\sigma}}_e-\bar{\bm{\sigma}}_e^*)
\end{equation}
Here, $\bar{\bm{\eta}}$ is a vector of Lagrange multipliers enforcing the equilibrium constraints, $\bar{\textbf{B}}_e$ is a matrix of interpolation functions prescribed by the finite element discretization, $\bar{\bm{f}}$ denotes the vector of nodal forces, $w_{e}$ denotes the integration weight and $m$ is the total number of integration points. Note that $\bar{\mathbb{C}}$ is a user-defined symmetric matrix to scale stress and strain to a similar magnitude, which does not represent a material behavior.

We take all possible variations of \cref{Eq01} considering the compatibility constraints $\bar{\bm{\varepsilon}}_{e}=\bar{\textbf{B}}_{e}\bar{\bm{u}}$, resulting in the following linear equations
\begin{subequations}\label{DD_linear}
\begin{align}
& \sum_{e=1}^{m} w_{e} \bar{\textbf{B}}_{e}^\text{T} \bar{\mathbb{C}}\bar{\textbf{B}}_{e}\bar{\bm{u}}=\sum_{e=1}^{m} w_{e} \bar{\textbf{B}}_{e}^\text{T} \bar{\mathbb{C}}\bar{\bm{\varepsilon}}_{e}^{*} \\
& \sum_{e=1}^m w_e \bar{\textbf{B}}_{e}^\text{T} \bar{\mathbb{C}} \bar{\textbf{B}}_{e} \bar{\bm{\eta}}=\bar{\bm{f}}-\sum_{e=1}^m w_e \bar{\textbf{B}}_{e}^\text{T} \bar{\bm{\sigma}}_e^*
\end{align}
\end{subequations}
The data-driven computing starts by randomly selecting data $\bar{\bm{z}}_e^*$ from the database $\bar{\mathscr{D}}$ for each integration point, followed by an iteration involving two steps. The first step is a mapping from $\bar{\bm{z}}_e^*$ to $\bar{\bm{z}}_e$ by solving the linear problem in \cref{DD_linear}. The second step is a reverse mapping from $\bar{\bm{z}}_e$ to $\bar{\bm{z}}_e^*$ via nearest-neighbor search, aiming at finding to find the closest material data $\bar{\bm{z}}_e^*$ in the database for each admissible data $\bar{\bm{z}}_e$
\begin{equation}
	\bar{\mathscr{F}}_e(\bar{\bm{z}}_e,\bar{\bm{z}}_e^*)
	\leq
	\bar{\mathscr{F}}_e(\bar{\bm{z}}_e,\bar{\bm{z}}'_e),\quad\forall \bar{\bm{z}}'_e \in \bar{\mathscr{D}}
\end{equation}
This requires distance calculations between $\bar{\bm{z}}_e$ and all the data $\bar{\bm{z}}'_e \in \bar{\mathscr{D}}$, then the optimal data $\bar{\bm{z}}_e^*$ is set to be the data $\bar{\bm{z}}'_e$ corresponding to the minimum distance. The overall iteration stops when the distance between $\bar{\bm{z}}_e$ to $\bar{\bm{z}}_e^*$ is minimized. 


It should be emphasized that the nearest-neighbor search consumes a tremendous computational cost of distance calculation for a high-dimensional and high-density database. For a nearest-neighbor search involving a database with $N$ data, each of dimension $D$, the complexity of the brute-force search on a classical computer is $O(ND)$. Related works show that efficient data structures can reduce the complexity \cite{korzeniowski2021multi,friedman1977algorithm,bahmani2021kd,kdtree2, kdtree1} without significant loss of accuracy in data-driven computing. For instance, a $k$-d tree approach can substantially reduce the number of distance calculations, resulting in a complexity of approximately $O(\log(N)D)$ \cite{friedman1977algorithm}. To further reduce the computational cost, the $k$-d tree will be used in this work, equipped with quantum computing for distance calculation introduced in the next section.


\subsection{Distance calculation via quantum computing}\label{dis_quan}



{\color{black}In this paper, we integrate two quantum algorithms for distance calculation into data-driven computational homogenization. For simplicity, we denote the scaled admissible data $\bar{\bm{z}}_e$ as the vector $\bm{V}$ and the scaled material data $\bar{\bm{z}}'_e$ as the vector $\bm{V}'$, both having dimension $D$. The distance in \cref{Eq02} can then be rewritten as the squared Euclidean distance $d$
\begin{equation}\label{dis}
d = \left|\bm{V} - \bm{V}'\right|^2
\end{equation} 
The two quantum algorithms for computing $d$ are introduced below.

\begin{figure}[t]
	\centering
	\includegraphics[width=13cm]{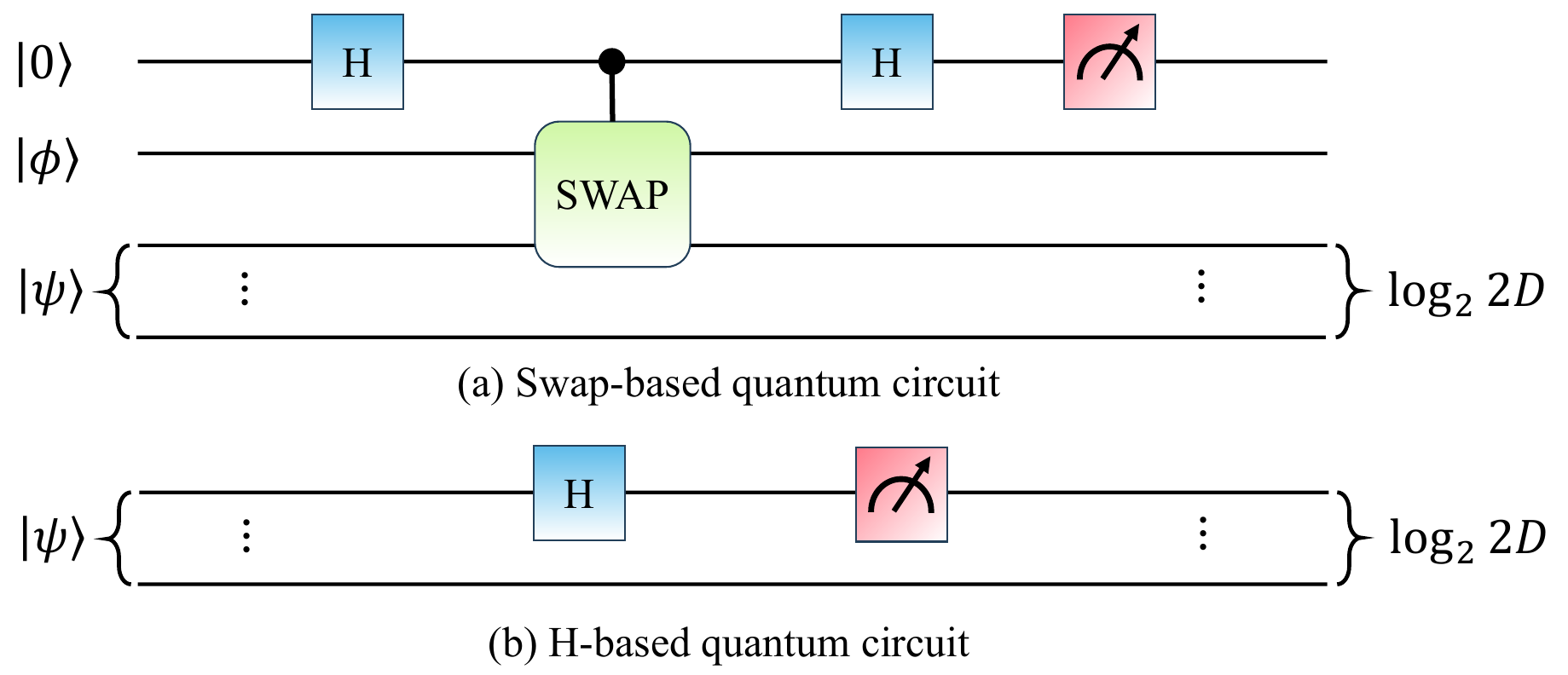}
	\caption{Quantum circuits of the two quantum algorithms.}
	\label{fig: sketch_circuit}
\end{figure}

The first is the Swap test-based quantum algorithm (Swap-based) \cite{xu2023quantum}, with its quantum circuit shown in \cref{fig: sketch_circuit} (a). Interested readers can refer to \ref{sec: swap} for further details. Here, we introduce a new quantum algorithm based on a single Hadamard gate (H-based), which is built upon the Hadamard test quantum algorithm \cite{moradi2022clinical,blank2020quantum}.
 Compared to the Swap-based algorithm, it requires fewer qubits and quantum gates, exhibiting noteworthy advantages. The quantum circuit of the H-based algorithm is shown in \cref{fig: sketch_circuit} (b).}  It only needs to prepare the quantum state $|\psi\rangle$ via qRAM in time $O(\log D)$
\begin{equation}\label{state_psi}
|\psi\rangle=\frac{1}{\sqrt{2}}(|0\rangle|\bm{V}\rangle+|1\rangle|\bm{V}'\rangle)
\end{equation}
Then only one Hadamard gate is applied on the first qubit of $|\psi\rangle$, indicating a time complexity of $O(1)$. The probability of the measurement result of the first qubit being $|0\rangle$ is
\begin{equation}\label{H-p}
\begin{split}
p_h = \frac{1}{2} + \frac{1}{2}{\langle \bm V  | \bm V' \rangle }
\end{split}
\end{equation}
where the information of the inner product $\langle \bm V  | \bm V' \rangle$ is contained in the probability. Then, by using the law of cosines, the distance is calculated by
\begin{equation}\label{H-d}
d = Z-2| \bm{V} || \bm{V}'|\left(2{p}_{h}-1\right)
\end{equation}
Similar to the Swap-based algorithm, multiple measurements of the first qubit are also needed to estimate $p_h$, i.e., if $n_0$ of the $n_m$ measurements are $|0\rangle$, then $p_h$ is estimated by $\hat{p}_h=n_0/n_m$. By substituting $\hat{p}_h$ in \cref{H-d}, we have
\begin{equation}\label{H-d2}
\hat{d} = Z-2| \bm{V} || \bm{V}'|\left(2\frac{n_0}{n_m}-1\right)
\end{equation}
where the estimated distance $\hat{d}$ for $d$ is obtained.

{\color{black}Complexity analysis for the two quantum algorithm is provided in \ref{sec: complexity analysis}, which shows that both algorithms can reduce the computational complexity of distance calculation from $O(D)$ to $O(\log D)$. By using these quantum algorithms, the complexity of the nearest-neighbor search in data-driven method is reduced to $O(D + N\log(D))$ \cite{xu2023quantum}. The performance of the Swap-based and H-based algorithms in distance calculation is evaluated in \cref{sec:Validation of distance}.}
In addition, the simulations in this work are conducted with the $k$-d tree structure \cite{friedman1977algorithm} to reduce the number of queries at the same time, resulting in a more favorable complexity $O(D + \log(N)\log(D))$. 
In the next section, the zero-noise extrapolation technique (referred to as ZNE) is employed to mitigate the error caused by quantum hardware noise, thereby improving the accuracy of the quantum algorithms in distance calculation. For clarity in subsequent sections, the probabilities derived from Swap-based {\color{black}(see \ref{sec: swap})} and H-based algorithms will no longer be distinguished. Instead, the true and estimated probabilities are denoted as $p$ and $\hat{p}$, respectively.
 
\subsection{Zero-noise extrapolation}\label{sec:ZNE}
While quantum computing has demonstrated a remarkable exponential reduction in computational complexity, the accuracy of current devices of quantum computing is influenced by hardware noise \cite{kandala2019error}, leading to challenges in extracting precise and effective information. Specifically, the estimated $\hat{p}$ will be biased from the true $p$, causing an error in the calculated distance. Here, the zero-noise extrapolation is applied to improve the accuracy of the distance calculation.
It offers a way to mitigate quantum hardware noise without the need for additional qubit resources and detailed knowledge of quantum processors.
The main idea of ZNE is to intentionally scale up the hardware noise so as to obtain the estimated $\hat{p}$ with different noise levels. Then the obtained results are processed on a classical computer to extrapolate the noiseless value. In the following, we will introduce the details of ZNE, which can be implemented in two steps: noise-scaling and extrapolation \cite{giurgica2020digital}.

\begin{figure}[!htbp]
	\centering
	\includegraphics[width=11cm]{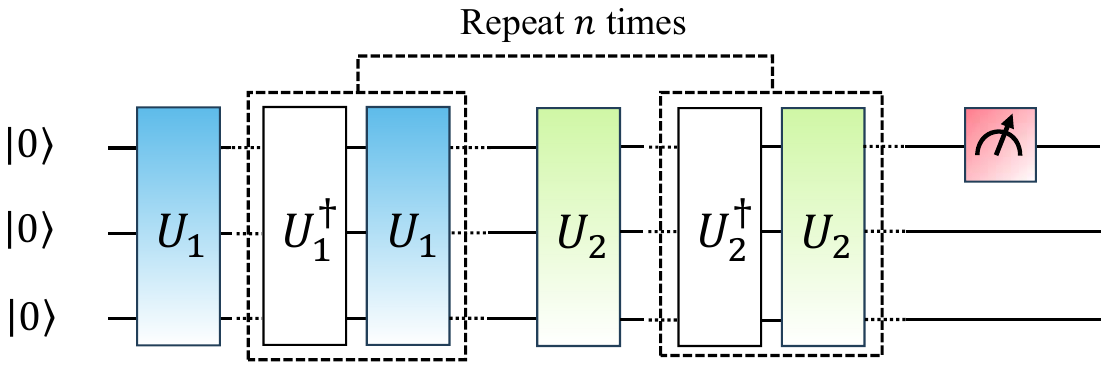}
	\caption{Sketch of the folding circuit.}
	\label{fig:folding circuit}
\end{figure}

\subsubsection{Scaling up the noise}
The first step aims to scale up the hardware noise while still allowing the quantum circuit to perform distance calculation. 
To achieve this, 
the quantum circuit of the Swap-based or the H-based algorithm requires to be modified by folding each quantum gate in it. Here, we use $U$ to represent an arbitrary quantum gate in the circuit, e.g., the Hadamard gate in the circuit of the Swap-based algorithm. Then the folding of the gate is represented by
\begin{equation}
U \longrightarrow U(U^{\dagger}U)^i
\end{equation}
where $U^{\dagger}$ is the conjugate transpose of $U$ and $i$ is a positive integer representing the number of foldings. If all the gates are noiseless, the gate folding has no effect on the final output since $U^{\dagger}U$ is equal to the identity operator \cite{Nielsen}. However, on a quantum computer with hardware noise, the gate folding will lead to an increasing noise with a factor $\lambda = 1+2i$, which corresponds to the total number of gates in the folding circuit. Hence, we can construct a folding circuit by repeating each gate of the quantum algorithm (see \cref{fig:folding circuit}), and perform measurements on the quantum circuit to get the estimated probability $\hat{p}$ with higher noise. 
By controlling the number of foldings $i$ from $0$ to a maximum folding number $n$ to build a set of quantum circuits, we can get the estimated $\bm{\hat{p}}= \{\hat{p}_0, \hat{p}_1, \dots, \hat{p}_{n}\}$ corresponding to a set of noise scale factors $\boldsymbol{\lambda}=\{1, 3, \dots, 1+2n\}$.

In the following, we will illustrate the influence of gate folding on the estimated $\hat{p}$ with noisy quantum computing, which requires deriving the relation between $\hat{p}$ and $\lambda$. 
We would like to mention that, for all the numerical examples in this work, the quantum hardware noise model including depolarizing channel and thermal relaxation is employed, and their parameters are obtained from IBM's real device (see \ref{append:noise_model}). Here, for simplicity in demonstration, only the single-qubit depolarizing channel is employed in the following derivations.
The density matrix is used to describe the mixed quantum state concerning decoherence. 
Firstly, we consider a single qubit density matrix $\rho$, whose evolution with an noiseless gate $U$ is expressed as%
\begin{align}
\text{Noiseless gate:}& \nonumber\\
&\rho \stackrel{U}{\longrightarrow}  U  \rho U^{\dagger}
\end{align}
Then we consider the hardware noise of depolarizing channel. If we define that the probability of the qubit being depolarized is $q$, then a noisy gate $U$ applying to the density matrix can be described as
\begin{align}
\text{Noisy gate $(\lambda = 1)$:}& \nonumber\\
\rho \stackrel{U}{\longrightarrow}& (1-\frac{3}{4}q) U \rho U^{\dagger}+\frac{q}{4}\sum_{j=1}^3(UG_j)\rho(UG_j)^{\dagger},~G=\{X, Y, Z\}
\label{eq:noise gate}
\end{align}
where $G = \{X,Y,Z\}$ are three Pauli gates, respectively to a rotation around the $x$, $y$ and $z$ axes of the Bloch sphere by $\pi$ radians \cite{Nielsen}. 
For simplicity, the second term of \cref{eq:noise gate} is set to $\displaystyle\rho^{\star}(q) = \frac{q}{4}\sum_{j=1}^3(UG_j)\rho(UG_j)^{\dagger} $, which represents a mixed state. Then \cref{eq:noise gate} can be rewritten as
\begin{align}
\text{Noisy gate $(\lambda = 1)$:}&\nonumber\\
\rho \stackrel{U}{\longrightarrow} &(1-\frac{3}{4}q) U \rho U^{\dagger}+\rho^{\star}(q)\label{eq:12}
\end{align}
Now we consider folding the gate once, which means we need to apply two additional gates $U^{\dagger}$ and $U$. First, we apply the noisy gate $U^{\dagger}$
\allowdisplaybreaks
\begin{align}
\text{Noisy gate $(\lambda=2)$:} &\nonumber\\
\quad \rho \xrightarrow{UU^{\dagger}} & (1-\frac{3}{4}q)U^{\dagger}((1-\frac{3}{4}q)U \rho U^{\dagger}+\rho^{\star}(q))U+\nonumber\\
& \frac{q}{4}\sum_{j=1}^3(UG_j)^{\dagger}((1-\frac{3}{4}q) U \rho U^{\dagger}+\rho^{\star}(q))(UG_j)\nonumber\\
=& (1-\frac{3}{4}q)^2\rho + (1-\frac{3}{4}q)U^{\dagger}\rho^{\star}(q)U+\nonumber\\
& \frac{q}{4}\sum_{j=1}^3(UG_j)^{\dagger}((1-\frac{3}{4}q) U \rho U^{\dagger}+\rho^{\star}(q))(UG_j)\nonumber\\
=&(1-\frac{3}{4}q)^2\rho + \rho^{\star}(q,q^2)
\label{eq:noise folding1}
\end{align}
where $\rho^{\star}(q,q^2)$ is a density matrix related to the linear combination of $q$ and $q^2$. Then we consider another noisy gate $U$ to finish the folding
\allowdisplaybreaks
\begin{align}
\text{Noisy gate $(\lambda = 3)$:} &\nonumber\\
\quad \rho \xrightarrow{UU^{\dagger}U} & (1-\frac{3}{4}q)U((1-\frac{3}{4}q)^2 \rho+\rho^{\star}(q,q^2))U^{\dagger}+\nonumber\\
& \frac{q}{4}\sum_{j=1}^3(UG_j)((1-\frac{3}{4}q)^2 \rho+\rho^{\star}(q,q^2))(UG_j)^{\dagger}\nonumber\\
=& (1-\frac{3}{4}q)^3U\rho U^{\dagger} + (1-\frac{3}{4}q)U\rho^{\star}(q,q^2)U^{\dagger}+\nonumber\\
& \frac{q}{4}\sum_{j=1}^3(UG_j)((1-\frac{3}{4}q)^2 \rho+\rho^{\star}(q,q^2))(UG_j)^{\dagger}\nonumber\\
=& (1-\frac{3}{4}q)^3U\rho U^{\dagger} + \rho^{\star}(q,q^2,q^3)
\label{eq:noise folding2}
\end{align}
One can see that after folding once, $\rho^{\star}$ is related to a higher order of $q^3$. Furthermore, if we consider folding $n$ times, the derivation of the folding can be generalized as
\begin{align}
\text{Noisy gate $(\lambda=1+2n)$:} &\nonumber\\
\rho \xrightarrow{U(U^{\dagger}U)^n} & (1-\frac{3}{4}q)^{\lambda}U\rho U^{\dagger} + \rho^{\star}(q,q^2,\dots, q^{\lambda})
\label{eq:noise folding3}
\end{align}
\cref{eq:noise folding3} corresponds to the folding of a single-qubit gate. Moreover, we generalize the gate folding for an arbitrary circuit, which is assumed to contain $m_s$ single-qubit gates and $m_t$ two-qubit gates. 
In total, there will be $m_d=m_s+2m_t$ times of single depolarizing acting on the qubits
\begin{align}
\text{Noisy circuit $(\lambda=1+2n)$:} &\nonumber\\
\rho \xrightarrow{\text{circuit}} & (1-\frac{3}{4}q)^{m_d\lambda}U\rho U^{\dagger} + \rho^{\star}(q^1,q^2,\dots, q^{m_d\lambda})
\label{eq:noise folding4}
\end{align}
where the density matrix $\rho$ is considered as the outer product of the initial quantum state $|0\phi\psi\rangle\langle0\phi\psi|$.
To obtain the probability of the first qubit measured to be $|0\rangle$, we need to sum the related diagonal elements of the final density matrix \cite{Nielsen} in \cref{eq:noise folding4} 
\begin{equation}
\hat{p}(\lambda) = (1-\frac{3}{4}q)^{m_d\lambda} p + \sum_{j=1}^{m_d\lambda}p^{\star}_{j}q^{j}
\label{eq:noise folding5}
\end{equation}
where $\hat{p}$ is the probability concerning hardware noise, $p$ is the true probability computed from the density matrix $U\rho U^{\dagger}$, and $p^{\star}_{j}$ are values computed from $\rho^{\star}(q^1,q^2,\dots, q^{m_d\lambda})$. 
One notes that the coefficient of the first term $(1-\frac{3}{4}q)^{m_d\lambda}$ will approach zero with the increasing noise scale factor $\lambda$, which means the information of $p$ gradually loses. 
In addition, 
this theoretical derivation is different from that in the work of Giurgica-Tiron et al. \cite{giurgica2020digital}. The reason is that their derivation is under the assumption of global depolarizing noise affecting all qubits, while this derivation considers depolarizing noise for each single qubit. %

\begin{figure}[!htbp]
	\centering
	\includegraphics[width=9cm]{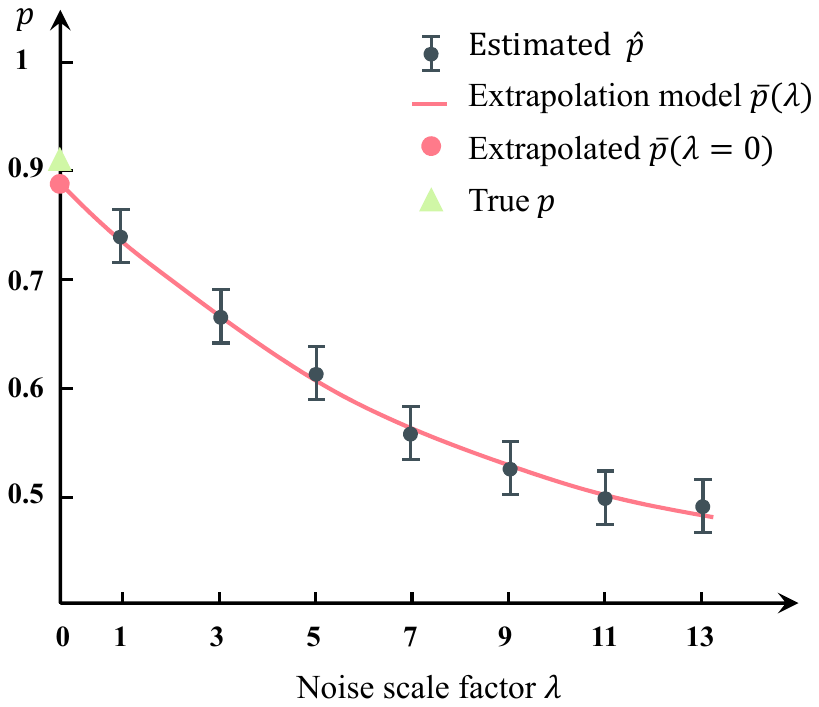}
	\caption{Sketch of the extrapolation method.}
	\label{fig:extrapolation}
\end{figure}

\subsubsection{Extrapolating the noiseless probability}
The second step of ZNE intends to extrapolate the noiseless probability based on the estimated $\bm{\hat{p}}= \{\hat{p}_0, \hat{p}_1, \dots, \hat{p}_{n}\}$ under different noise scale factors $\boldsymbol{\lambda}=\{1, 3, \dots, 1+2n\}$, which is processed on a classical computer.
As shown in \cref{fig:extrapolation}, this extrapolation requires a model $\bar{p}(\lambda)$ to characterize the relationship between $\hat{p}$ and $\lambda$. After fitting this model with $\bm{\hat{p}}$ and $\boldsymbol{\lambda}$, the  noiseless value is predicted as $\bar{p}(\lambda=0)$. One straightforward thought is to directly derive the relation between $\hat{p}$ and $\lambda$ considering various noise models shown in \ref{append:noise_model}, and then the derived relation is employed as the extrapolation model $\bar{p}(\lambda)$. However, even for the relation in \cref{eq:noise folding5} that only considers the single-qubit depolarizing channel, it is already hard to fit since the number of terms in  $\sum_{j=1}^{m_d\lambda}p^{\star}_{j}q^{j}$ changes depending on the value of $\lambda$, let alone more noise models are considered.
Another approach is to assume a relatively concise model between the estimated $\hat{p}$ and the noise scale factor $\lambda$, which is the core concept of ZNE.
Here, we present four extrapolation models, i.e., linear extrapolation, quadratic extrapolation, exponential extrapolation and Richardson extrapolation, which are commonly used in ZNE \cite{giurgica2020digital}
\begin{equation}
\begin{aligned}
& \text{Linear extrapolation~:} &&\bar{p}^{\text {linear }}(\lambda)=c_0+c_1 \lambda \\
& \text{Quadratic extrapolation~:} &&\bar{p}^{\text {quad }}(\lambda)=c_0+c_1 \lambda +c_2 \lambda^2 \\
& \text{Exponential extrapolation~:}&& \bar{p}^{\exp }(\lambda)=c_0+c_1 e^{-c_2 \lambda} \\
& \text{Richardson extrapolation~:} &&\bar{p}^{\text {Rich }}(\lambda)=c_0+c_1 \lambda+\ldots + c_{n} \lambda ^{n} 
\end{aligned}
\label{eq:extrapolation_method}
\end{equation}
where $c_0, c_1,\dots , c_{n}$ are the fitting coefficients. The linear, quadratic and Richardson extrapolations all belong to polynomial models. The difference lies in that the highest order of linear and quadratic extrapolations are fixed, while the highest order of the Richardson $n$ is related to the number of elements in the estimated probabilities $\bm{\hat{p}}= \{\hat{p}_0, \hat{p}_1, \dots, \hat{p}_{n}\}$. 
For example, the Richardson extrapolation will degenerate into the linear one when $n = 1$, and will degenerate into the quadratic one when $n = 2$. The coefficients of these models are determined via the least squares fitting, i.e., finds optimal coefficients that minimize the sum of squared residuals $\sum_{i=0}^{n}(\hat{p}_i-\bar{p}(\lambda=1+2i))^2$. Then the noiseless value can be predicted by $\bar{p}(\lambda=0)$. Finally, we substitute $\bar{p}(\lambda=0)$ into the distance expression of the Swap-based algorithm \cref{dis_final2} or the H-based algorithm \cref{H-d2}, resulting in the mitigated distance $\bar{d}$.

The performance of these extrapolation models will be validated in \cref{sec:Validation of distance}. We would like to mention that ZNE is used to handle the hardware noise associated with quantum gates, instead of the estimation error arising from the statical estimation of $\hat p$, which is presented by the grey intervals in \cref{fig:extrapolation}. However, the estimation error may affect the performance of ZNE, since a larger value of $n_m$ leads to lower estimation error, resulting in a higher accuracy of the estimated $\hat{p}$.  The influence of $n_m$ on error mitigation will be discussed later in \cref{sec:Validation of distance}. In addition, the flowchart of the distance calculation via error-mitigated quantum computing is detailed in \cref{alg:distance calculation}.

\begin{algorithm}
\caption{Distance calculation via error-mitigated quantum computing}\label{alg:distance calculation}
\begin{algorithmic}
\Require Vectors $\bm V$, $\bm V^\prime$; $|\bm V|^2$; $|\bm V^\prime|^2$; Number of measurements $n_m$; \\\hspace{1.6cm}Maximum folding number $n$

\State \textbf{Step 1: Noise-scaling}
\For{$i = 0 \text{~to~} n $}
\State  Fold each quantum gate of the Swap-based circuit or the H-based circuit $i$ times 
\State  Save the noise scale factor $\lambda=1+2i$ to the set $\bm\lambda$
\For{$t = 1 \text{~to~} n_s $}
\State  Run the circuit and measure the first qubit on a quantum computer
\EndFor
\State Set $n_0$ as the number of 0s in the measurement results
\State  Estimate the probability $\hat{p}=n_0/n_s$ and save it to the set $\hat{\bm p}$
\EndFor
\State \textbf{Step 2: Extrapolation}
\State Choose an extrapolation model $\bar{p}(\lambda)$
\State Fit $\bar{p}(\lambda)$ with $\hat{\bm p}$ and $\bm \lambda$
\State Extrapolate the noiseless value $\bar{p}(\lambda=0)$
\State Compute the mitigated distance $\bar{d}$ using \cref{dis_final2} or \cref{H-d2}

\end{algorithmic}
\end{algorithm}

%
%

\section{Validation}\label{sec:Validation}
In this section, numerical tests are carried out to validate the proposed data-driven method with error-mitigated quantum computing (referred to as mitigated qDD). First, we evaluate the performance of the Swap-based and the H-based algorithms in distance calculation considering hardware noise. Second, ZNE is employed in the two quantum algorithms and the performance of different extrapolation models on error mitigation are compared. Finally, a roof truss structure is used to validate the effectiveness of the mitigated qDD. 
Since the availability of real quantum computers is currently limited, 
the numerical tests in this paper are all carried out based on the quantum computer simulator Qiskit developed by IBM \cite{Qiskit}. To reflect the actual operation of a quantum computer, the quantum gates in circuits are transpiled into a standard set of basic gates $\{I, X, SX, RZ, ECR\}$, which are compatible with real quantum devices. Considering that qRAM is not yet available \cite{aaronson2015read, biamonte2017quantum}, a series of quantum gates are used to prepare the quantum states \cite{shende2005synthesis} in the two algorithms for distance calculation, and hardware noise is also considered in these gates and mitigated by ZNE. The folding circuit in ZNE is constructed based on the transpiled circuit. 
To simulate the hardware noise, each gate in the circuit is associated with a depolarizing noise followed by a thermal relaxation noise, and the noise parameters are obtained from the IBM's real device $ibm\_osaka$, as shown in \ref{append:noise_model}.  In addition, we use the normal approximation to accelerate the statistical estimation of $\hat{p}$ on the quantum computer simulator, as shown in \ref{sec: Normal}.


\subsection{Evaluation of the two quantum algorithms}\label{sec:Validation of distance}
The performance of the Swap-based and the H-based quantum algorithms in distance calculation is first evaluated, where the results with and without hardware noise are both presented. We randomly sample 1000  vector pairs $(\bm{V}, \bm{V}')$ in a 6-dimensional space, such that their distances $d=|\bm{V}-\bm{V}'|^2$ obey the uniform distribution. The Swap-based algorithm requires 6 qubits and has a circuit depth of around 100, whereas the H-based algorithm necessitates 4 qubits with a circuit depth of around 70. For both algorithms, the number of measurements $n_m$ is set to $10^{4}$.

\begin{figure}[t]
\centering
\subfloat[]{\label{fig:Actual_against_predict}
\includegraphics[width=0.48\textwidth]{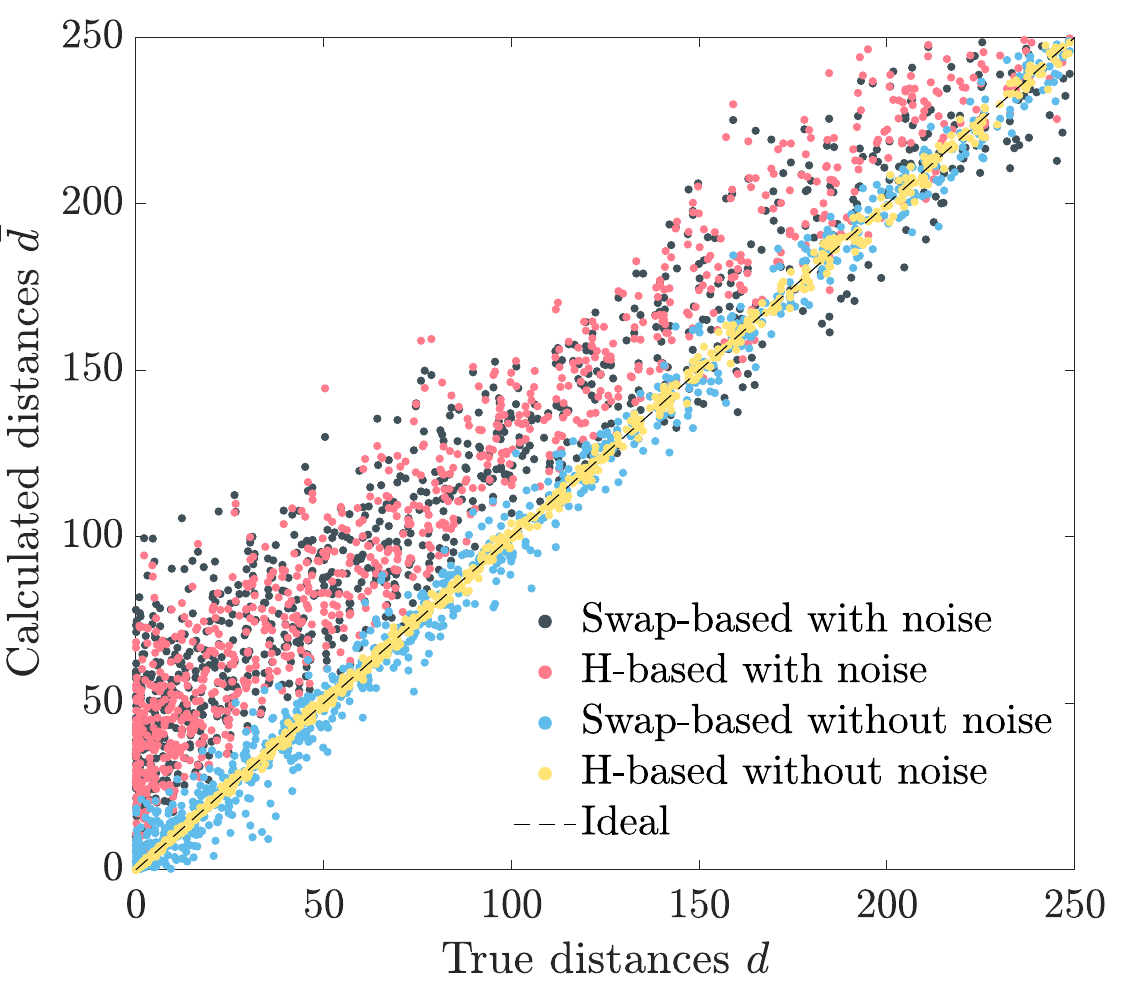}}
\subfloat[]{
\centering
\label{fig: distance error}
\includegraphics[width=0.48\textwidth]{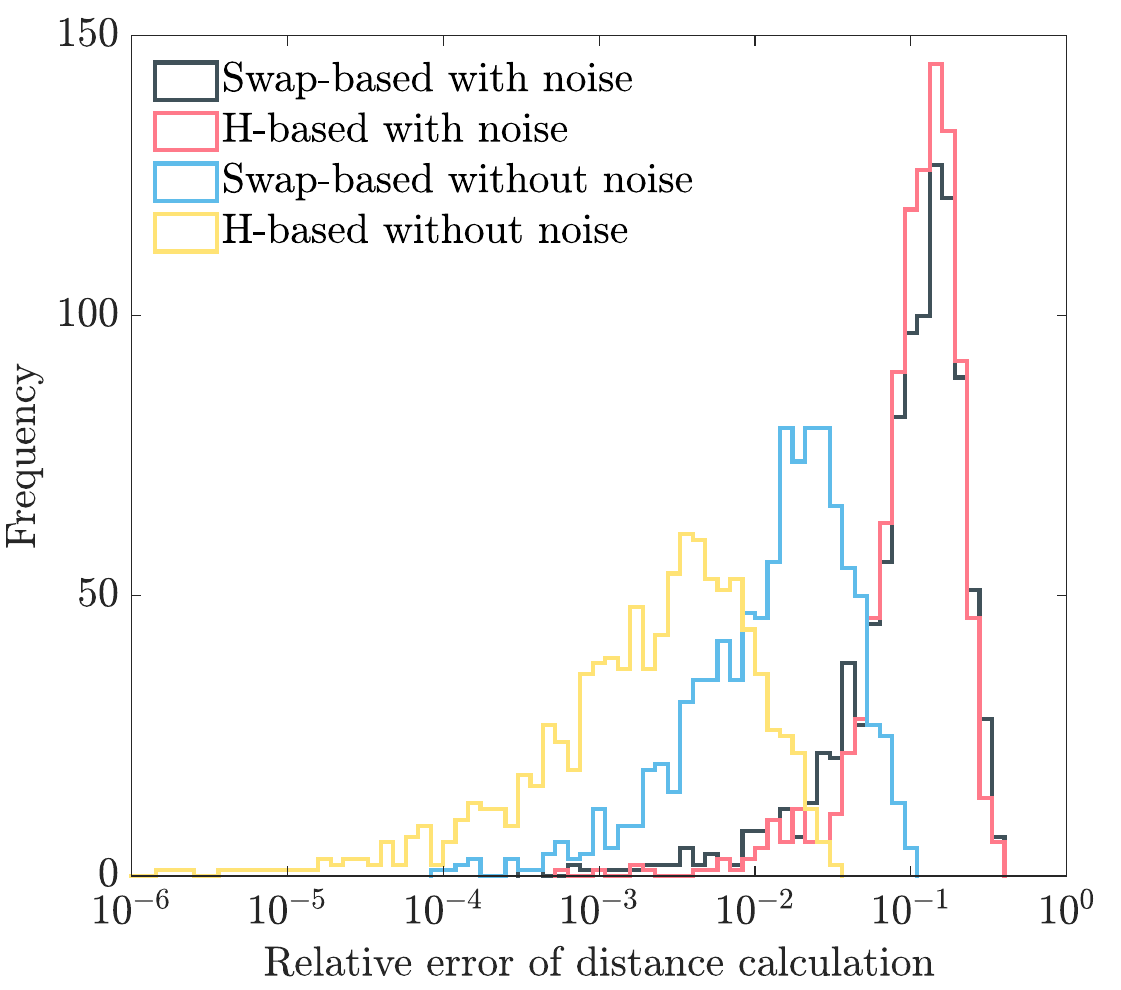}}
\caption{Results of the calculated distances with and without quantum hardware noise. (a) True distances versus calculated distances. (b) Frequency of the relative errors. }
\label{fig: result of dist}
\end{figure}

\cref{fig: result of dist} (a) represents the true distances versus calculated distances from the 1000 vector pairs. As a reference, the dashed line marked with `Ideal' is drawn according to the relation $\bar{d}=d$, meaning results closer to the dashed line exhibit higher accuracy. \cref{fig: result of dist} (b) shows the distribution of relative errors ${|\bar{d}-d|}/{d_{\max}}$, where $d_{\max}$ refers to the maximum value of $d$. In summary, the results considering hardware noise exhibit a larger error for both algorithms, compared to the results without noise. Specifically, the relative errors for both algorithms increase from around $10^{-2}$ to around $10^{-1}$ due to hardware noise. If the hardware noise is not considered, the H-based algorithm exhibits smaller errors than the Swap-based one, which is consistent with the error analysis in \cref{sec: complexity analysis}. However, when hardware noise is taken into account, the error distributions of the two algorithms are almost the same. This appears counterintuitive, as the Swap-based algorithm should have a lower accuracy since it employs more quantum gates with hardware noise. We believe the reason is that both algorithms require $O(D)$ layers of quantum gates to prepare the quantum state $ \left|\psi\right\rangle$ \cite{shende2005synthesis}.
As the dimension $ D $ increases, the depth of the circuit for state preparation increases linearly, while the depth of the rest of the circuit remains unchanged. Therefore, state preparation gradually takes a significant portion of the entire circuits for both algorithms. In this case, the hardware noise introduced during the state preparation becomes the predominant source of error in distance calculations, resulting in a similar accuracy for the two algorithms.

\begin{figure}[t]
	\centering
	\includegraphics[width=9cm]{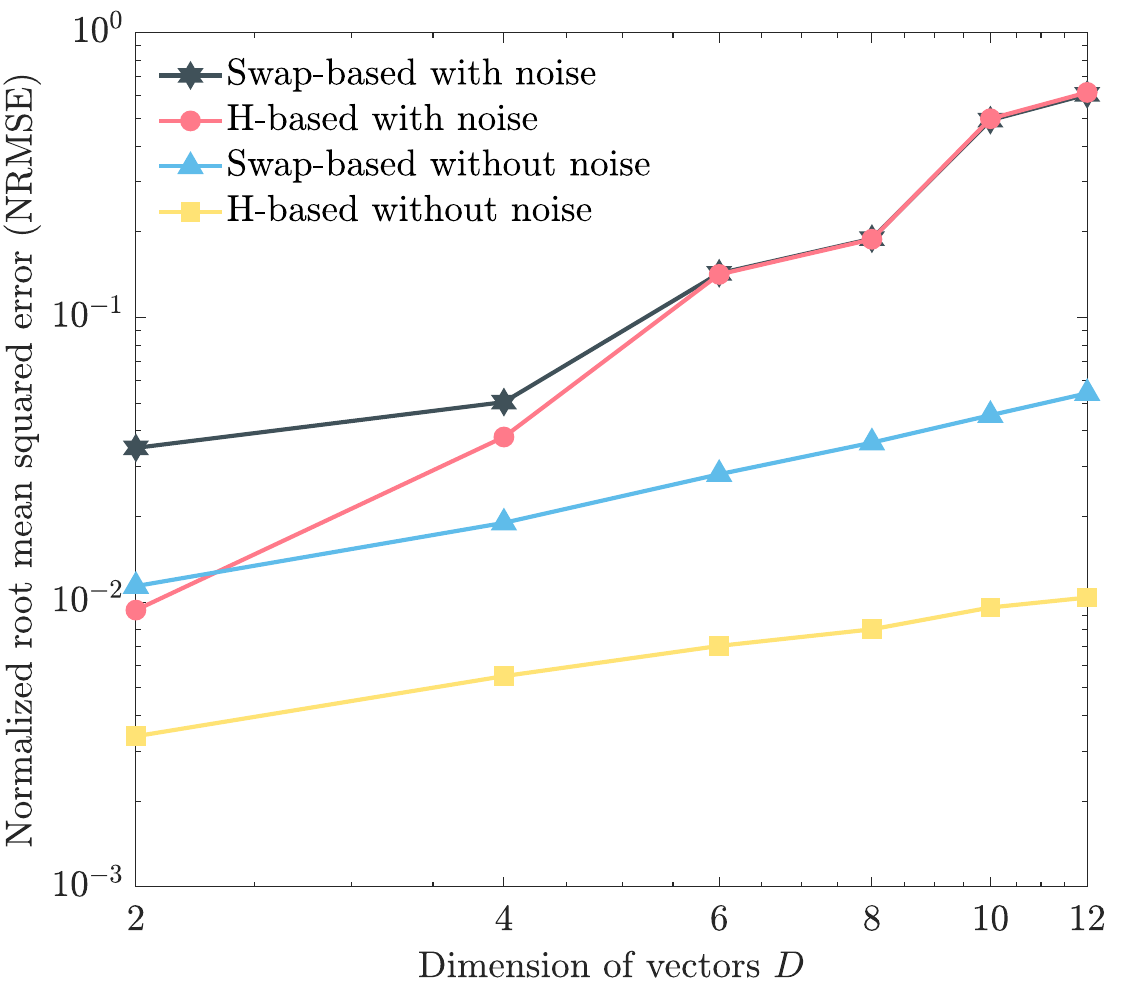}
	\caption{Dimension of vectors $D$ versus the normalized root mean squared error (NRMSE) of the calculated distances.}
	\label{fig:RMSE_dimension}
\end{figure}

To confirm the above explanation, the relationship between the dimension of vectors $D$ and the accuracy of distance calculations is presented in
\cref{fig:RMSE_dimension}. The normalized root mean squared error (NRMSE), defined as $\sqrt{{\sum(\bar{d}-d)^2}/({1000d_{\max}^2})}$, is used to reflect the average accuracy. As $D$ increases from 2 to 12, the portion of the state preparation in the entire circuit increase from $11/43~(26\%)$ to $212/244~(87\%)$ for the Swap-based algorithm, and from $10/13~(77\%)$ to $212/215~(99\%)$ for the H-based algorithm. This shows that as $D$ increases, the state preparation takes the majority of the proportion of the whole circuit. Correspondingly, if hardware noise is taken into account, the accuracy of the two quantum algorithms tends to be the same as shown in \cref{fig:RMSE_dimension}, which provides evidence for the above explanation.

In conclusion, the above analysis shows that hardware noise reduces the accuracy of distance calculations in both the Swap-based and H-based quantum algorithms. In the subsequent section, we explore the application of the ZNE, aiming to improve the accuracy of distance calculations.

\subsection{Performance of ZNE}\label{sec:Validation of ZNE}
\begin{figure}[t]
	\centering
	\includegraphics[width=9cm]{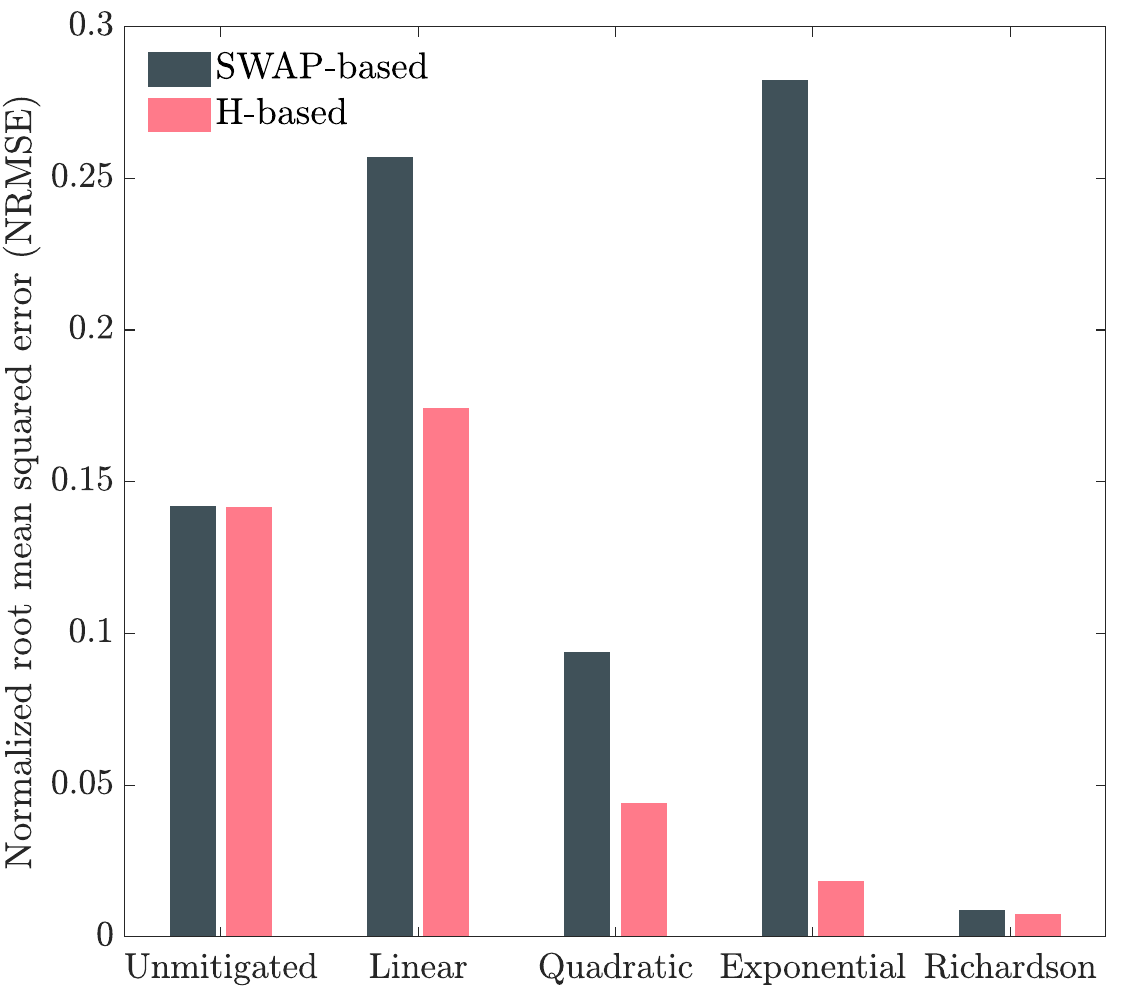}
	\caption{Normalized root mean squared error (NRMSE) of the calculated distances with ZNE.}
	\label{fig:RMSE_ZNE}
\end{figure}

In this section, the performance of ZNE in error mitigation for the two quantum algorithms is investigated. We use the linear, quadratic, exponential, and Richardson extrapolation models to predict noiseless probability $\bar{p}(\lambda=0)$ and then calculate the mitigated distance $\bar{d}$. First, the number of measurements $n_m$ is set to $10^8$ and the maximum folding number $n$ is set to 6, which means $\bm \lambda=\{1,3,\dots, 13\}$ and $\hat{\bm{p}}=\{\hat{p}_0,\hat{p}_1,\dots,\hat{p}_6\}$ are used to fit each extrapolation model $\bar{p}(\lambda)$. As shown in \cref{fig:RMSE_ZNE}, the Richardson extrapolation model demonstrates superior performance across all models. It can reduce the NRMSE from 14.18\% to 0.87\% for the Swap-based algorithm, and from 14.14\% to 0.74\% for the H-based algorithm, resulting in a similar level of accuracy for the two algorithms. This validates the effectiveness of ZNE in improving the accuracy of distance calculation. We believe that the Richardson extrapolation model outperforms the other three models primarily due to its inclusion of more fitting coefficients, which enhance its fitting capabilities.

\begin{figure}[htbp]
	\centering
	\subfloat[]{
		\label{fig: RMSE_shotsnum_SWAP}
		\includegraphics[width=0.47\linewidth]{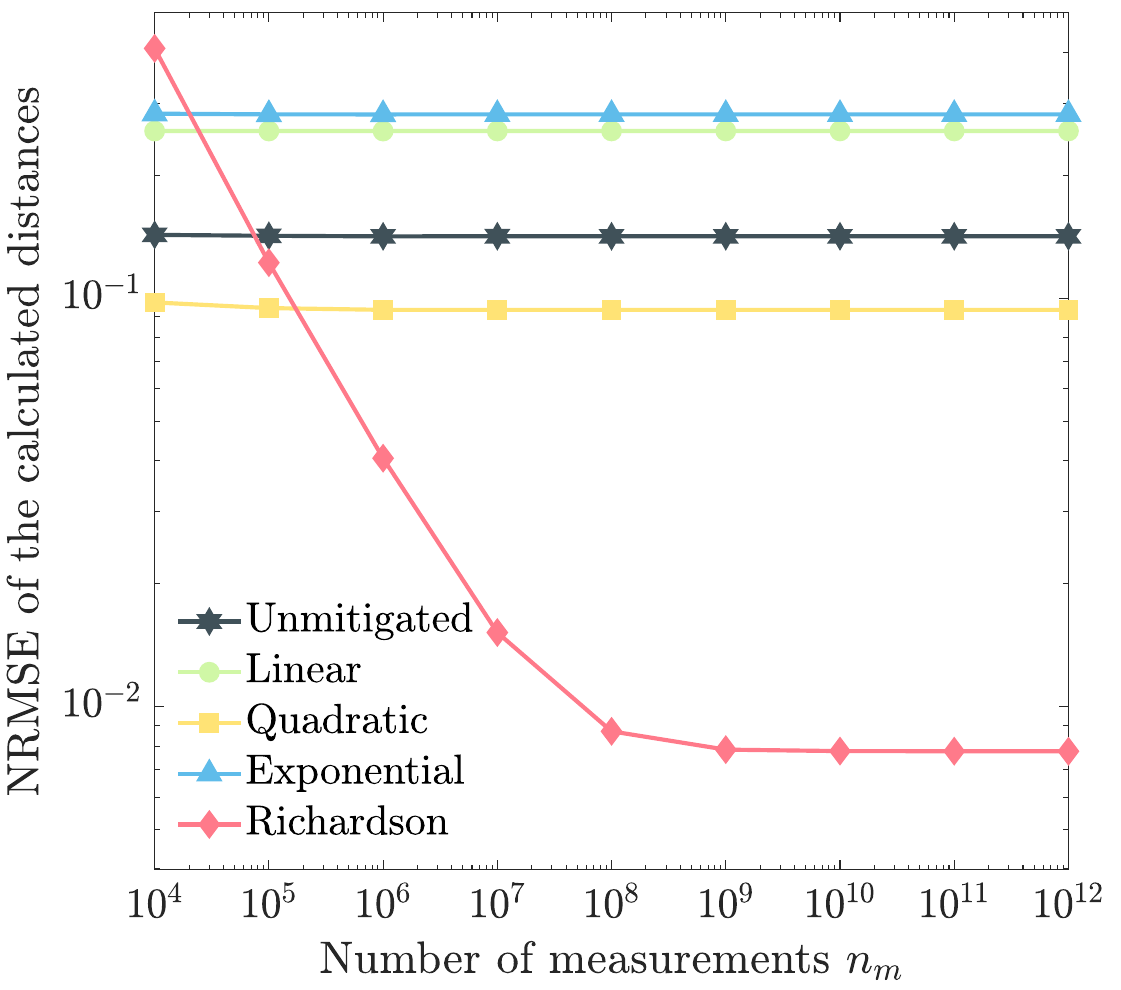}}
	\subfloat[]{
		\label{fig: RMSE_shotsnum_H}
		\includegraphics[width=0.47\linewidth]{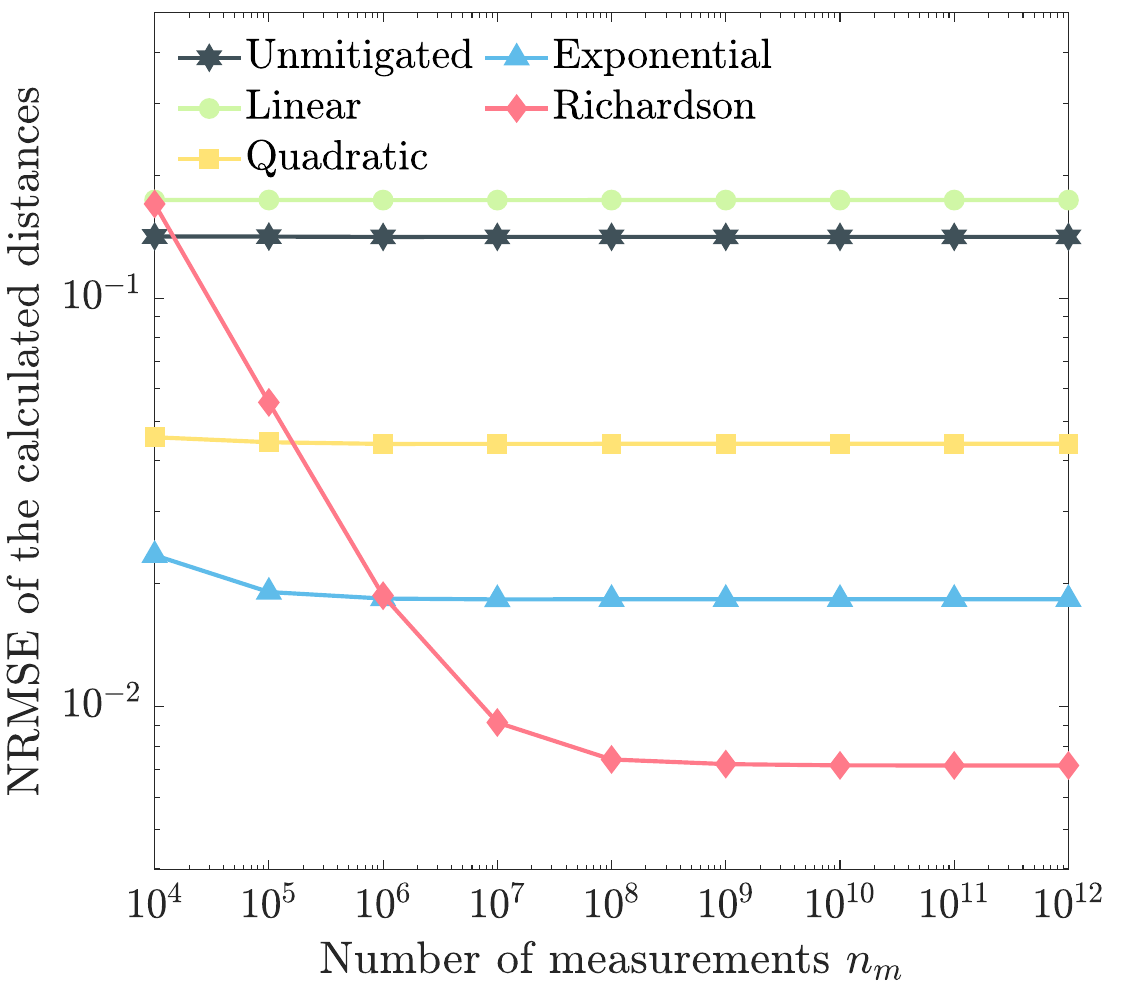}}
	\caption{Influence of the number of measurements $n_m$ on the (a) Swap-based and (b) H-based algorithms with ZNE.}
	\label{fig: param_optim_nm}
\end{figure}

Moreover, we investigate the effect of $n_m$ on the performance of ZNE, where a larger value of $n_m$  leads to a lower estimation error of $\hat{\bm p}$, as mentioned in \cref{sec: complexity analysis}. 
\cref{fig: param_optim_nm} (a) and \cref{fig: param_optim_nm} (b) respectively present the influence of $n_m$ on the NRMSE of the Swap-based and the H-based algorithm, where the maximum folding number $n$ is set to 6. For both the Swap-based and the H-based algorithms, the performance of linear, quadratic, and exponential extrapolations are almost unaffected by $n_m$, while the performance of Richardson extrapolation shows a strong correlation with respect to $n_m$, i.e., its accuracy is improved with an increasing $n_m$ and tends to be converged when $n_m$ is about $10^{10}$.
The reason is that Richardson extrapolation has more fitting coefficients than the other three models. On one hand, it leads to a higher ability to represent a complex relationship between $\hat{p}$ and $\lambda$. On the other hand, 
it is more sensitive to the estimation error. This is consistent with the results in \cite{kandala2019error}, where it is found that the Richardson extrapolation is sensitive to the variance of the unmitigated measurements.
Therefore, the Richardson extrapolation can accurately predict the noiseless $\bar{p}(\lambda=0)$ and the mitigated $\bar{d}$ with a large $n_m$, and shows large errors with a small $n_m$. 

\begin{figure}[htbp]
	\centering
	\subfloat[]{
		\label{fig: RMSE_scale_factor_SWAP}
		\includegraphics[width=0.47\linewidth]{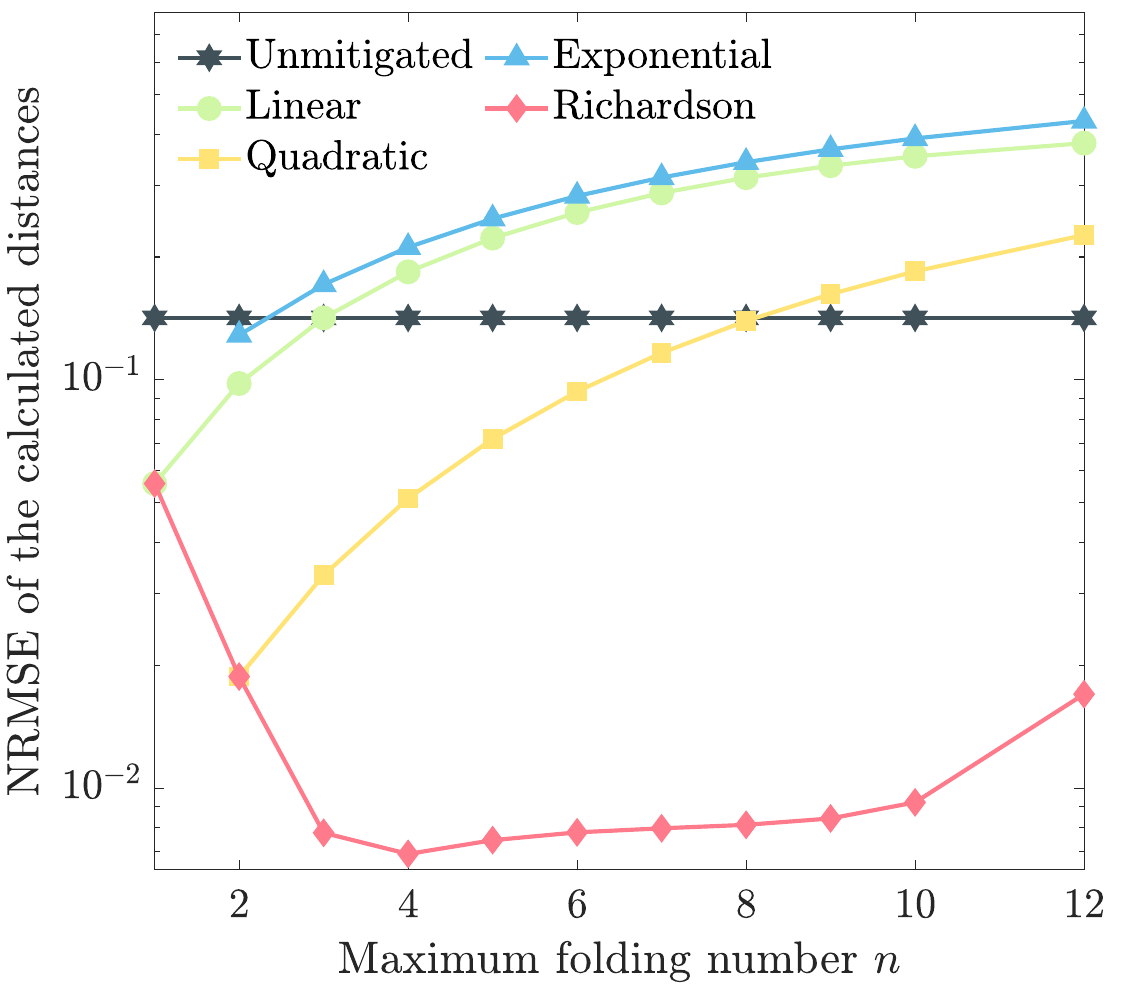}}
	\subfloat[]{
		\label{fig: RMSE_scale_factor_H}
		\includegraphics[width=0.47\linewidth]{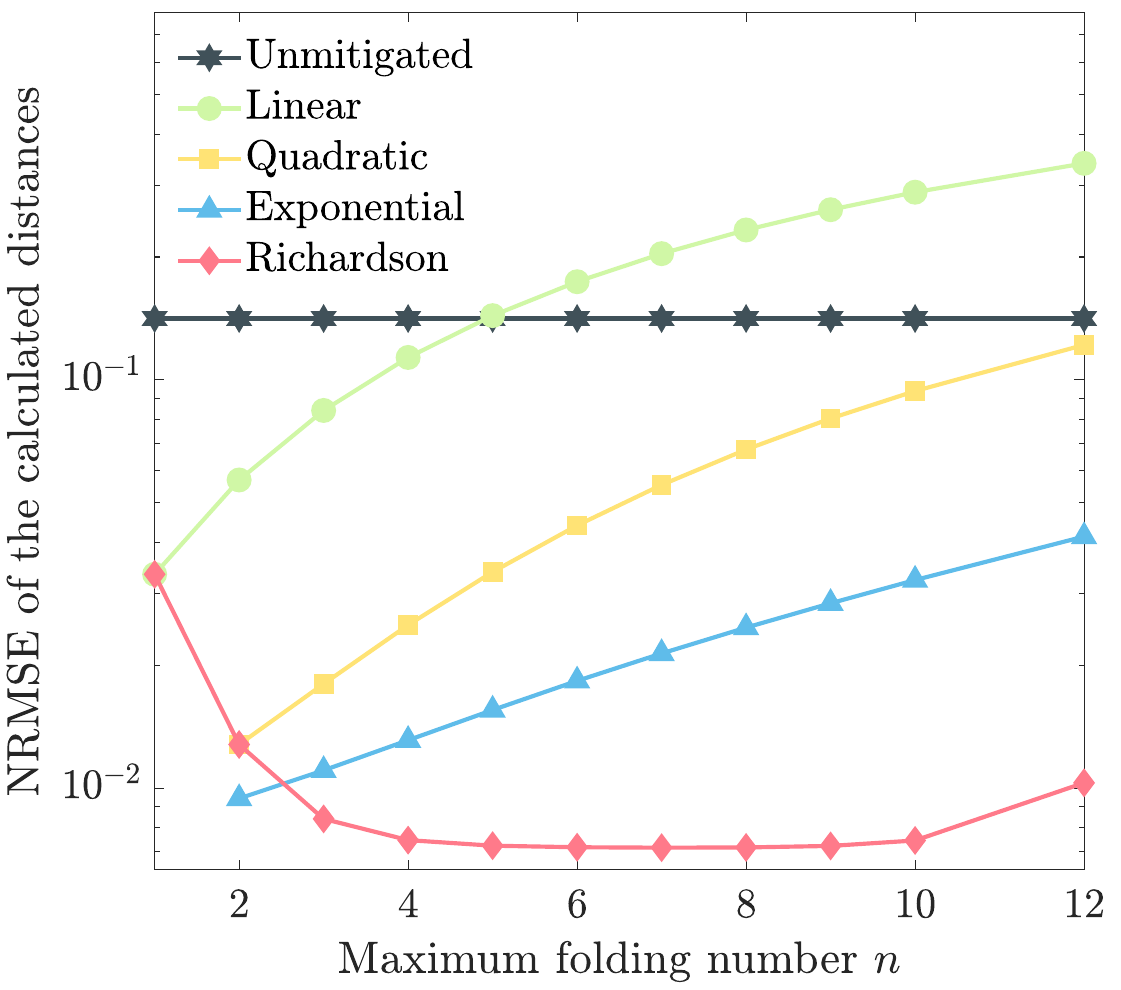}}\\
	
	\caption{Influence of maximum folding number $n$ on the (a) Swap-based and (b) H-based algorithms with ZNE.}
	\label{fig: param_optim_n}
\end{figure}
Furthermore, we investigate the influence of the maximum folding number $n$ on the performance of ZNE, where the number of measurements $n_m$ is set to $10^{10}$. As shown in \cref{fig: param_optim_n} (a) and \cref{fig: param_optim_n} (b), the performance of ZNE is sensitive to the maximum folding number $n$. The accuracy of the distance calculations with linear, quadratic, and exponential extrapolations decreases with the increasing $n$. Whereas the accuracy of the distance calculations with Richardson extrapolations reaches an optimum at $n=4$ for the Swap-based algorithm and $n\in [5,9]$ for the H-based algorithm. 
We would like to mention that the approaches for optimizing $n$ deserve future investigations \cite{krebsbach2022optimization}.

In conclusion, the numerical results demonstrate the effectiveness of ZNE in improving the accuracy of distance calculation with noisy quantum computing. Although the two algorithms can reach a similar level of accuracy with the help of ZNE, the H-based algorithm consumes fewer qubits and quantum gates compared to the Swap-based algorithm, exhibiting an advantage for the NISQ quantum computers. 
Hence, ZNE with the H-based algorithm is utilized in subsequent numerical simulations.

\vspace{2mm}
\noindent\textbf{Remark.} Compared to the H-based algorithm, the Swap-based algorithm has a unique ability to reduce the complexity of distance calculation between one data and the center of a cluster of data \cite{lloyd2013quantum}. However, the application of this ability is limited by the requirement of data normalization. One potential application is to construct the efficient $k$-means data structure for the material database \cite{kdtree2}, where the complexity of a standard $k$-means algorithm on a classical computer is $O(kND)$ for one iteration. However, to use the Swap-based algorithm, all the data are required to be normalized in the first place, which already takes $O(ND)$, eliminating the complexity advantage of quantum computing. Regarding another potential application in data search concerning $k$-means data structure \cite{kdtree2}, the standard way is to directly calculate the distance between the admissible point and the center of a cluster of data, where the latter is already computed offline in advance.
Therefore, there is no obvious advantage to using the Swap-based algorithm in this case either. When the center of a cluster of data can not be computed offline in advance, such as in the case of on-the-fly material data sampling \cite{karapiperis2021data}, the application of the Swap-based algorithm is once again prevented by the need for data normalization, since the data is newly generated.  In future research, investigating an efficient quantum algorithm for distance calculation that does not require data normalization could be valuable. In addition, we believe the unique advantage of the Swap-based algorithm in data-driven computing deserves future exploration.

\subsection{Roof truss}\label{validation}
\begin{figure}[tbp]
\centering
\includegraphics[width=9cm]{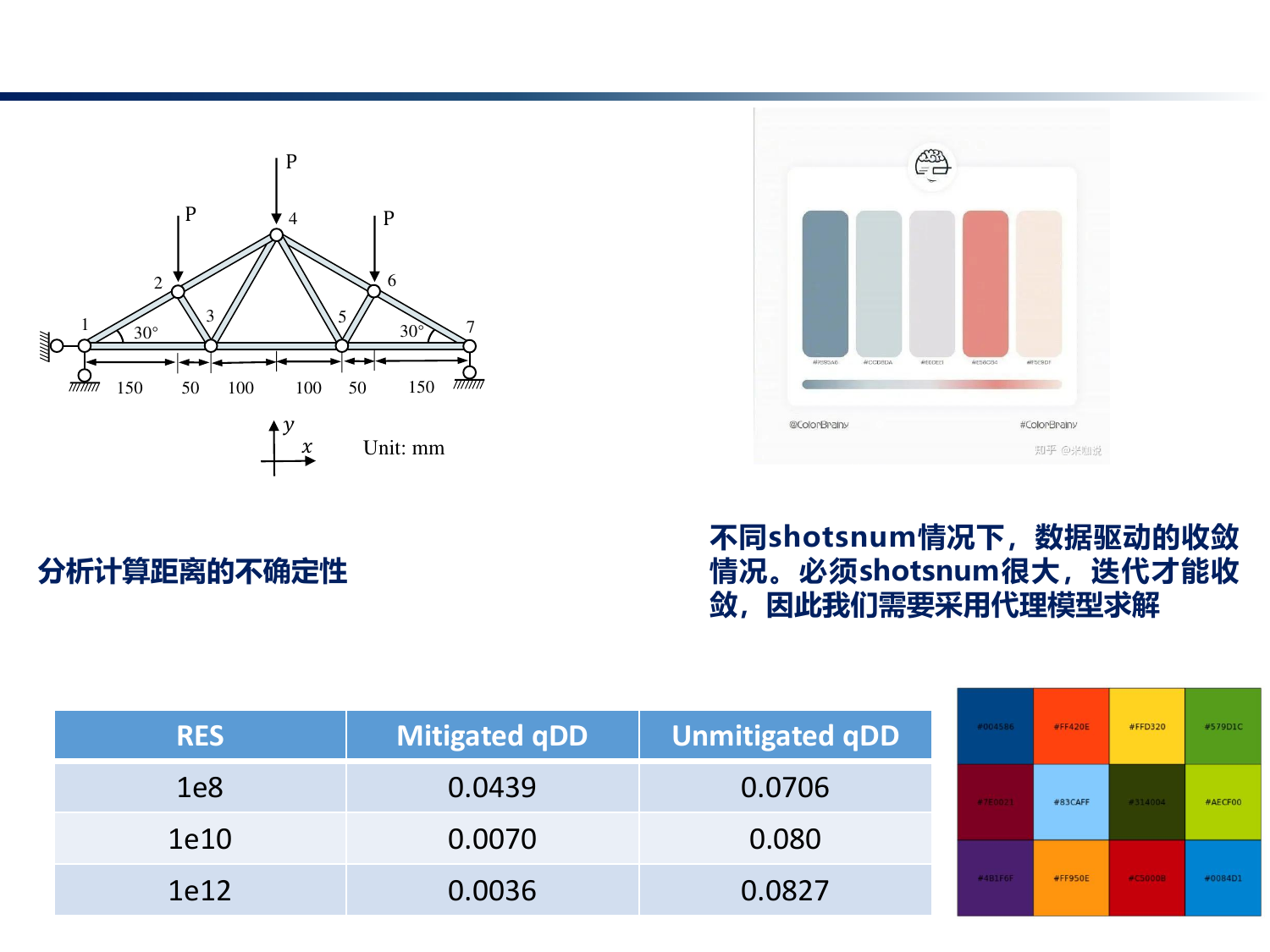}
\caption{Sketch of the roof truss.}
\label{sketch_truss_fig}
\end{figure}

In this section, a roof truss is considered to validate the performance of ZNE in data-driven simulation with noisy quantum computing. Here, the H-based algorithm is employed for distance calculation, and it necessitates 2 qubits with a circuit depth of around 15 without folding. The parameters $n$ and $n_m$ for ZNE are set to $5$ and $10^{10}$, respectively. The configuration of the roof truss is shown in \cref{sketch_truss_fig}. 
All the bars have the same cross-sectional area $A = 100~\text{mm}^2$. The displacement of node 1 is fixed in the $x$ and $y$ directions, while the node 7 is fixed in the $y$ direction. The nodes 2, 4 and 6 are under the loads $\rm{P} = 200~\text{N}$. 

The material of the roof is assumed to satisfy a Ramberg-Osgood material model, and the material database is collected by uniformly sampling the stress data in the range $\sigma\in[-6,6]~\text{MPa}$ and obtaining the corresponding strain data $\epsilon$ via the following relation 
\begin{equation}
    {\epsilon}=\frac{\sigma}{E}+\alpha \frac{\sigma}{E}\left(\frac{|\sigma|}{\sigma_0}\right)^{\beta-1}
\end{equation}
where $E=10000~\text{MPa}$ refers to Young's modulus, $\alpha=0.5$ the yield offset, $\sigma_0=5~\text{MPa}$ the yield stress and $\beta = 3$ the hardening exponent. In this way, a material database with $161$ data is generated. Furthermore, a $k$-d tree data structure is used to reduce the number of distance calculations in data-driven computing. The analytical solution is used as the reference result, and a root-mean-square (RMS) error of stress $\sigma_{\text{RMS}}$ is employed to evaluate the accuracy of solutions
\begin{equation}
\sigma_{\rm{RMS}}=\sqrt{\frac{\displaystyle\sum_{e=1}^{m}w_e(\sigma_e-\sigma_e^{ref})^2}{\displaystyle\sum_{e=1}^{m}w_e(\sigma_e^{ref})^2}}
\end{equation}

\begin{figure}[tbp]
	\centering
	\includegraphics[width=9cm]{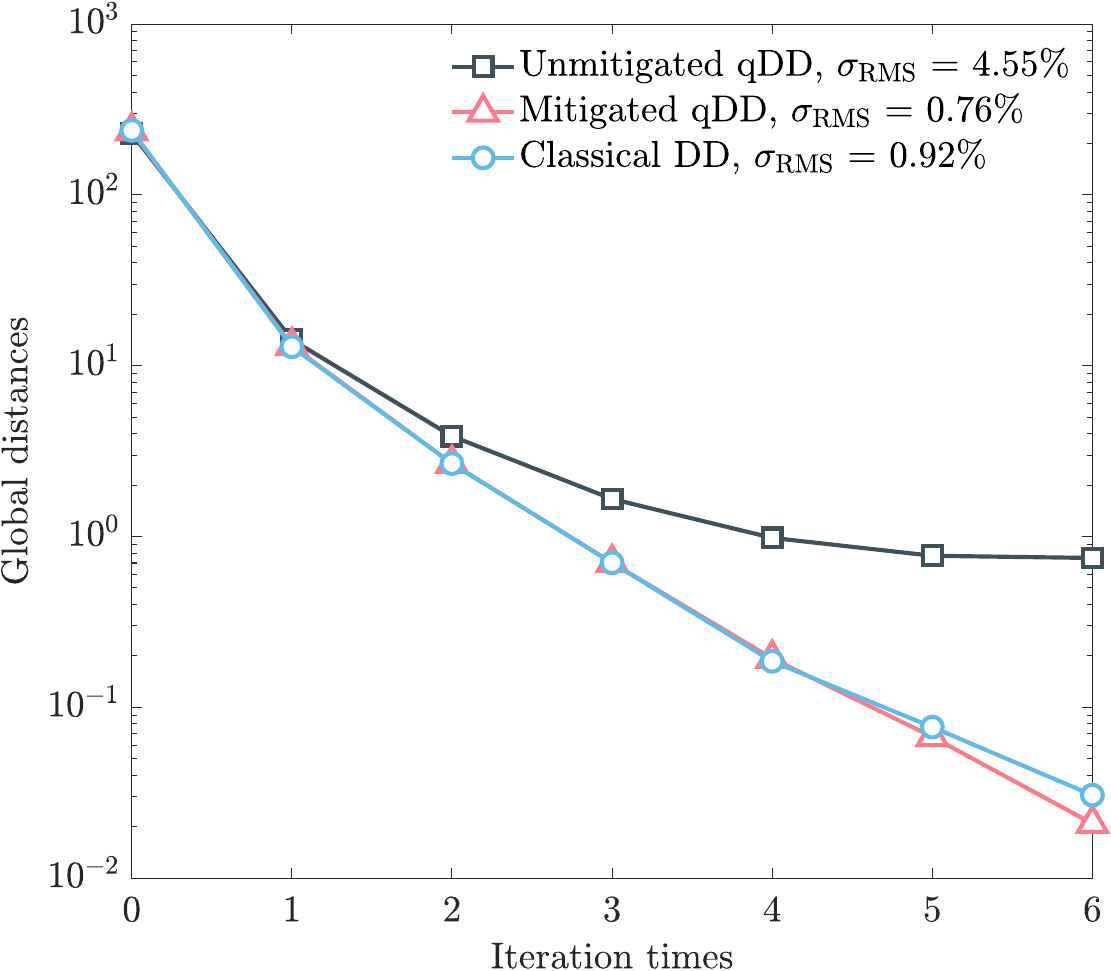}
	\caption{Global distances obtained by unmitigated qDD, mitigated qDD and classical DD.}
	\label{fig:ddc_iter_fig}
\end{figure}

Firstly, the performance of the data-driven methods based on classical computing (classical DD), noisy quantum computing (unmitigated qDD), and error-mitigated quantum computing (mitigated qDD) are evaluated. \cref{fig:ddc_iter_fig} presents the evolution of the global distance  $\bar{\mathscr F}=\frac{1}{2}\sum_{e=1}^{m}{w_e}\bar{\mathscr{F}}_e(\bar{\bm{z}}_e,\bar{\bm{z}}_e^*)$ during the data-driven computing process, which is defined as the sum of the distances at all the integral points \cite{Ortiz2016}. 
The mitigated qDD can accurately predict the stress of each bar ($\sigma_{\mathrm{RMS}}=0.76\%$), which is comparable to that of a classical DD ($\sigma_{\mathrm{RMS}}=0.92\%$). In comparison, although the unmitigated qDD also converged, the obtained result shows noticeable discrepancies from the reference solution ($\sigma_{\mathrm{RMS}}=4.55\%$). This means that ZNE can effectively improve the accuracy of qDD with noisy quantum computing. 

\begin{figure}[tbp]
	\centering
	\includegraphics[width=9cm]{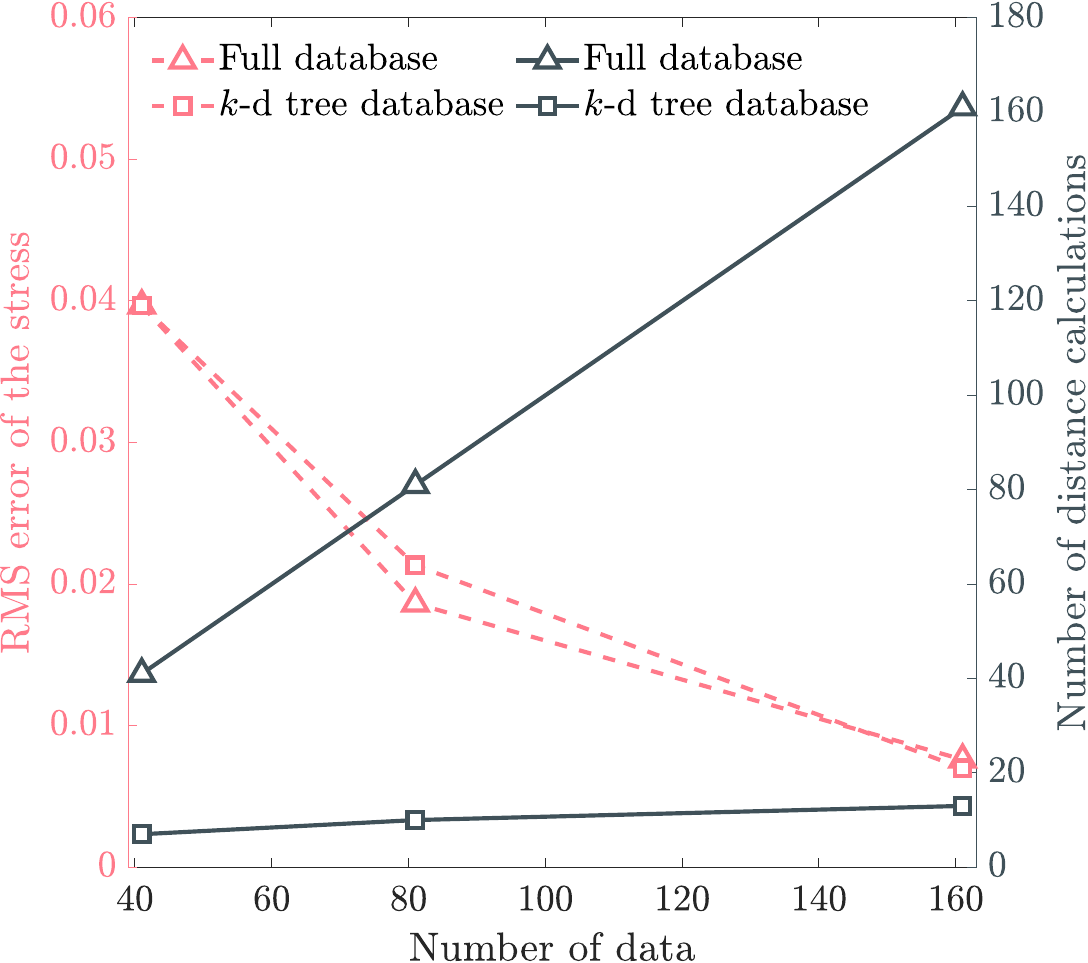}
	\caption{Performance of the $k$-d tree database compared with the full database.}
	\label{fig:databasecompare}
\end{figure}

Furthermore, the effectiveness of the $k$-d tree data structure in accelerating qDD is verified. \cref{fig:databasecompare} shows the RMS error of the stress and the average number of distance calculations per nearest-neighbor search versus the number of data in the database. Compared to the full database without data structure, the $k$-d tree database can achieve the same level of accuracy while significantly reducing the number of distance calculations. We would like to emphasize that the results show opportunities for combining a variety of efficient data structures \cite{kdtree2,kdtree1,bahmani2021kd,kuang2023data} with qDD, resulting in accelerations in both the number of data $N$ and the dimension $D$.

Finally, the influence of the number of measurements $n_m$ on the mitigated qDD is investigated. As shown in \cref{table:RMS_shots}, the value of $n_m$ will affect the accuracy of mitigated qDD but has almost no effect on the unmitigated qDD. This can be explained by the relation between $n_m$ and the accuracy of the distance calculation (see \cref{fig: param_optim_nm} (b)), where the mitigated distance with Richardson extrapolation is sensitive to $n_m$, while the unmitigated distance is almost unaffected by $n_m$. Since the main idea of data-driven computing is to find the material data $ \bar{\bm{z}}_e^{*}$ closest to the admissible state $\bar{\bm{z}}_e$, the accuracy of the calculated distance between $\bar{\bm{z}}_e^{*}$ and $\bar{\bm{z}}_e$ will directly influence the nearest-neighbor search, thereby affecting the performance of data-driven simulation.

In a word, error mitigation with ZNE can improve the performance of quantum computing in data-driven simulation.

\begin{table}[tbp]
\caption{Root-mean-square (RMS) error of the stress versus the number of measurements $n_m$ for the mitigated and unmitigated qDD. }
\centering
\renewcommand{\arraystretch}{1.5}
\renewcommand\tabcolsep{12.0pt}
\footnotesize{
\begin{tabular}{cccc}
\toprule
Classical DD &   \multicolumn{3}{c}{0.92\%}     \\
\hline
 & $n_s = 10^6$ &  $n_s = 10^{8}$ &  $n_s = 10^{10}$  \\
\cline{2-4}
Mitigated qDD &   4.79\%   & 1.56\%     &   0.76\%           \\ 
Unmitigated qDD     &   5.10\%   & 4.36\%     &   4.55\%      \\
\bottomrule
\end{tabular}}
\label{table:RMS_shots}
\end{table}

\section{Application \textcolor{black}{in a 2D case}: composite L-shaped beam}\label{sec:Application}

In this section, error mitigated quantum computing is applied to data-driven computational homogenization for the multiscale simulation of a \textcolor{black}{2D} composite L-shaped beam. The reference solutions are obtained by classical data-driven computational homogenization method (classical DD-FE$^2$), which is performed on classical computers. 
Here, the H-based quantum algorithm with ZNE is employed to reduce the computational complexity of data-driven multiscale simulation, and it necessitates 4 qubits with a circuit depth of around 70 without folding. The parameters $n$ and $n_m$ for ZNE are set to $5$ and $10^{10}$, respectively. Meanwhile, the $k$-d tree data structure is also employed to reduce the number of distance calculations, resulting in a more favorable computational complexity.

\begin{figure}[tbp]
\centering
\includegraphics[width=10cm]{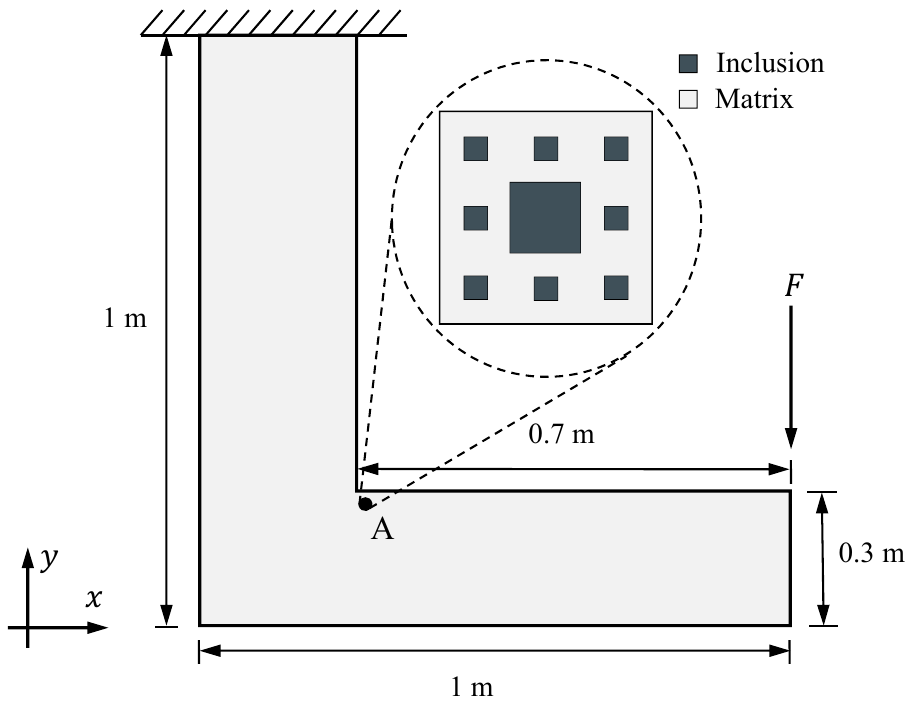}
\caption{Sketch of the L-shaped beam.}
\label{sketch_beam_fig}
\end{figure}

\begin{figure}[h]
\centering
\includegraphics[width=9cm]{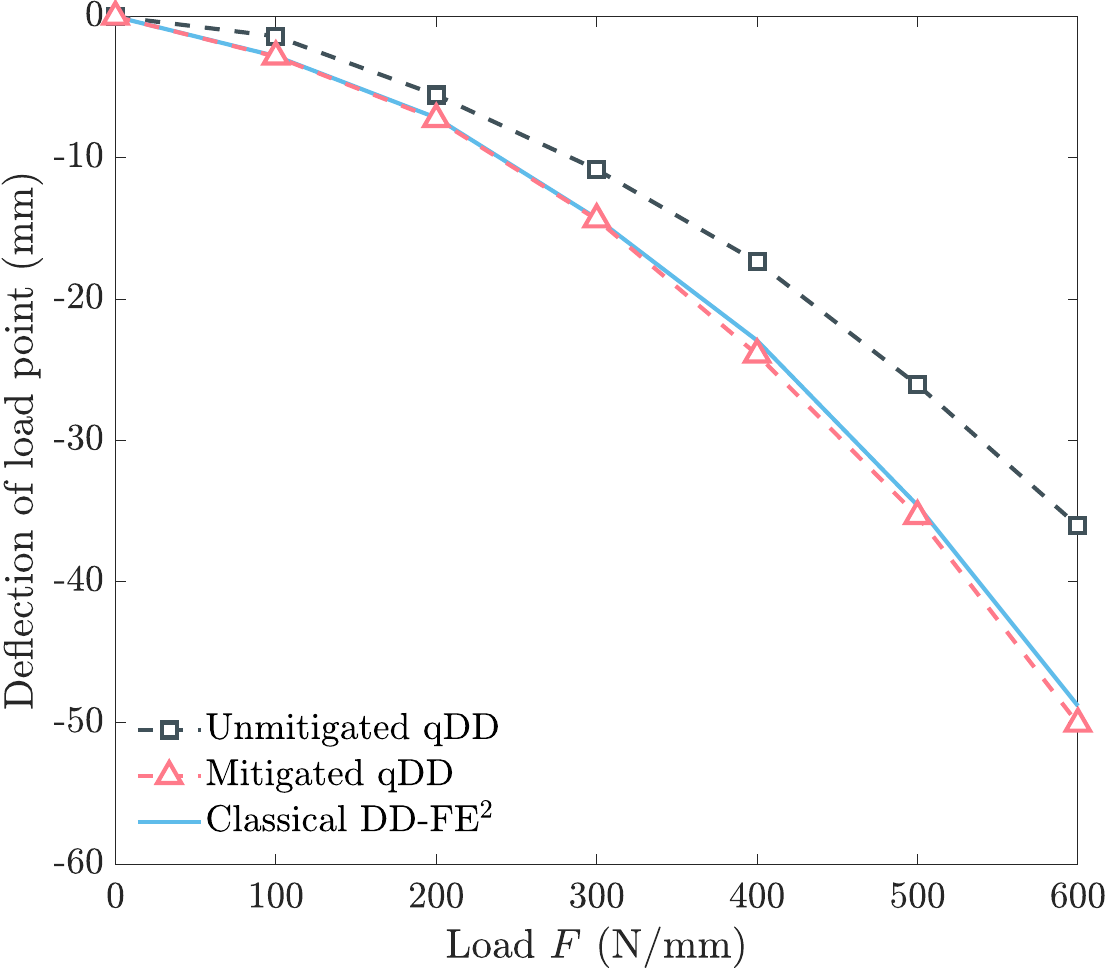}
\caption{Deflection of the load point versus the external load.}
\label{fig:LU_curve_L_beam}
\end{figure}

The macroscopic configuration of the L-shaped beam is shown in \cref{sketch_beam_fig}, as well as the RVE at the microscopic scale. The beam is subjected to a concentrated load $ F = 600~\mathrm{N/mm}$ on the right edge and a fixed constraint on the upper edge. The microstructure of the beam consists of the inclusion and the matrix, both of which are assumed to satisfy the Ramberg-Osgood constitutive relation 
\begin{equation}
E \boldsymbol{\varepsilon}=\left(1+\nu\right) \tilde{\boldsymbol{\sigma}}-\left(1-2 \nu\right) p \mathbf{I}+\frac{3}{2} \alpha\left(\frac{\sigma_{M}}{\sigma_0}\right)^{\beta-1} \tilde{\boldsymbol{\sigma}}
\end{equation}
where $\nu$ represents the  Poisson ratio, $\tilde{\boldsymbol{\sigma}}=\boldsymbol{\sigma}+p\mathbf{I}$ denote the stress deviator, $p=-\frac{1}{3}\boldsymbol{\sigma}:\mathbf{I}$ is the equivalent hydrostatic stress, $\sigma_{M}$ expresses the Mises equivalent stress. The material properties of the inclusion and the matrix are given in \cref{table:fractal}. To generate the material database, we uniformly sample the macroscopic strains in ranges $\bar{\varepsilon}_{xx}\in[-0.015,0.030]$, $\bar{\varepsilon}_{yy}\in[-0.025,0.040]$, and $\bar{\varepsilon}_{xy}\in[-0.015,0.010]$. Then the macroscopic stresses are computed through computational homogenization on the RVE, resulting in $100^3$ data.

\begin{table}[t]
\caption{Material parameters of the inclusion and the matrix.}
\centering
\renewcommand{\arraystretch}{1.5}
\renewcommand\tabcolsep{12.0pt}
\footnotesize{
\begin{tabular}{cccccc}
\toprule
 & $E$ (MPa) &  ${\sigma}_0$ (MPa) &  $\nu$ & $\alpha$  & $\beta$  \\
\hline 
Inclusion     &   $10^5$   & $100$     &   $0.3$   &     $0.5$      &        $3$           \\ 
Matrix     &   $10^4$   & $10$     &   $0.3$   &     $0.5$      &        $3$          \\ 
\bottomrule
\end{tabular}}
\label{table:fractal}
\end{table} 

\begin{figure}[tbp]
	\centering
	\subfloat[Mitigated qDD]{
		\label{fig: 2D_mitigated_qDD}
		\includegraphics[width=0.47\linewidth]{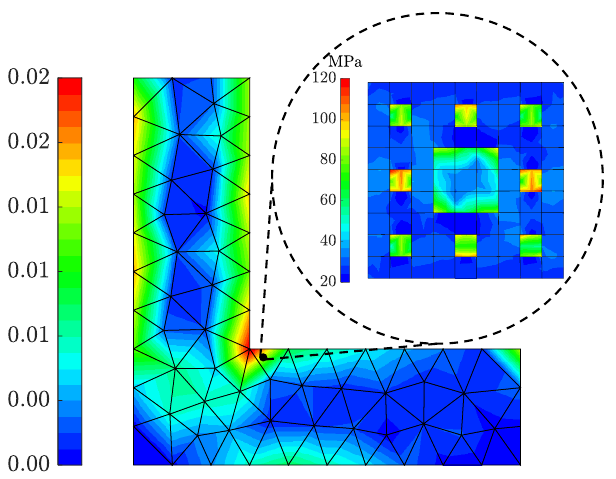}}
	\subfloat[Relative error of mitigated qDD]{
		\label{fig: 2D_mitigated_qDD_error}
		\includegraphics[width=0.47\linewidth]{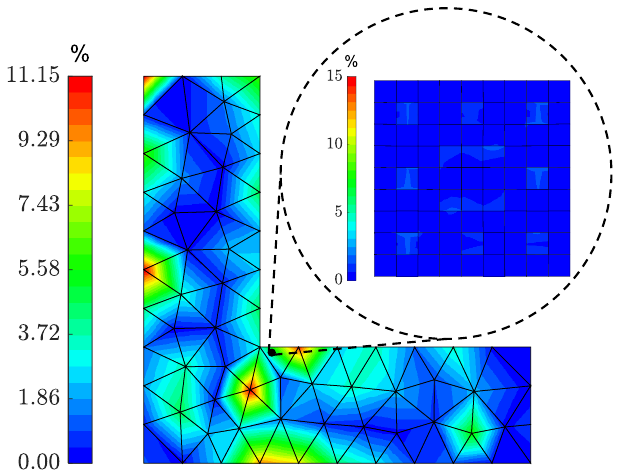}}\\
	\subfloat[Unmitigated qDD]{
		\label{fig: 2D_unmitigated_qDD}
		\includegraphics[width=0.47\linewidth]{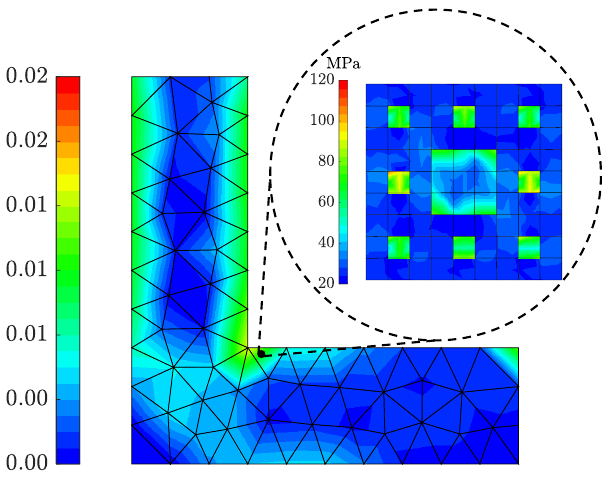}}
	\subfloat[Relative error of unmitigated qDD]{
		\label{fig: 2D_unmitigated_qDD_error}
		\includegraphics[width=0.47\linewidth]{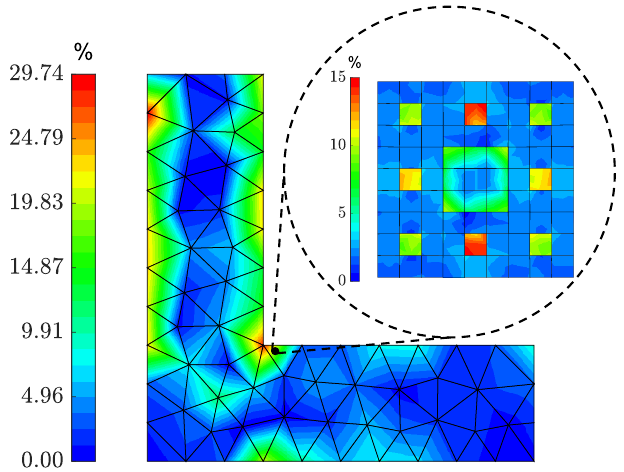}}\\
	
	\caption{Strain fields $\sqrt{3J_2}$ and the corresponding relative errors for the L-shaped beam, as well as the von Mises stress fields and the corresponding relative errors for the RVE at the integration point. The solutions of classical DD-FE$^2$ are used as a reference.}
	\label{fig: 2D_strain_field}
\end{figure}

For the computation at the macroscopic scale, the mitigated qDD and unmitigated qDD are applied. As a reference, the DD-FE$^2$ method with the classical computer (referred to as classical DD-FE$^2$) 
is used to solve the same problem.
\cref{fig:LU_curve_L_beam} shows the load-deflection curve of the load point. The mitigated qDD can accurately predict the deflection of the load point under various loads, while unmitigated qDD exhibits around 20\% relative errors.
\cref{fig: 2D_strain_field} shows the strain fields $\sqrt{3J_2}$ and the corresponding relative errors at the macroscopic scale under the load ${F}=600~\mathrm{N/mm}$, as well as the von Mises stress fields and the corresponding relative errors at the microscopic scale. Compared to the unmitigated qDD, the relative error of the stress field at the microscopic scale is notably reduced, and the maximum relative error at the macroscopic scale is reduced from $29.74\%$ to $11.15\%$. In a word, the mitigated qDD exhibits higher accuracy compared to the unmitigated qDD at both microscopic and macroscopic scales. 
This validates the effectiveness of error-mitigated quantum computing in data-driven multiscale simulation.

\section{\textcolor{black}{Application in a 3D case:  composite cylindrical shell}}\label{sec: 3d shell}
\textcolor{black}{In this section, we further evaluate the performance of the mitigated qDD with a 3D composite cylindrical shell, where a seven-parameter shell formulation \cite{buchter1994three,zahrouni1999computing,kuang2021computational} is employed. 
In this shell model, the displacement field $\bar{\mathbf{u}}(\theta_1, \theta_2, \theta_3)$ is represented by the mid-surface displacement $\bar{\mathbf{v}}(\theta_1, \theta_2)$ and the difference vector between the deformed and undeformed directors $\bar{\mathbf{w}}(\theta^1,\theta^2)$
\begin{equation}
\bar{\mathbf{u}}(\theta^1,\theta^2,\theta^3)=\bar{\mathbf{v}}(\theta^1,\theta^2)+\theta^3\bar{\mathbf{w}}(\theta^1,\theta^2)
\end{equation}
in which $(\theta_1, \theta_2, \theta_3)$ denote the curvilinear coordinates. To ensure that the strain component $\bar{\varepsilon}_{33}$ is distributed linearly along the thickness direction, which is essential for thin shells, an enhanced assumed strain $\bar{\bm\varepsilon}^a$\cite{simo1990class} is incorporated in the displacement-dependent compatible strains $\bar{\bm\varepsilon}^u$
\begin{equation}
\bar{\bm\varepsilon}=\bar{\bm\varepsilon}^u+\bar{\bm\varepsilon}^a\label{eq: enhanced assumed strain}
\end{equation}
in which $\bar{\bm\varepsilon}^a$ is linearly distributed across the thickness and only $\bar{\varepsilon}^a_{33}$ is existed. Based on the three-field Hu-Washizu functional, the potential energy of the shell with domain $\Omega$ is written as
\begin{equation}
\displaystyle
{\varPi} =\int_\Omega\{W(\bar{\bm\varepsilon})-\bar{\bm\sigma}:\left(\bar{\bm\varepsilon}-\bar{\bm\varepsilon}^u\right)\}\text{d}\Omega-\mathbf{P}_e(\bar{\mathbf{u}})
\label{eq:Hu-Washizu}
\end{equation}
where $W(\bar{\bm\varepsilon})$ denotes the strain energy and $\mathbf{P}_e(\bar{\mathbf{u}})$ is the potential energy of external forces. By substituting \cref{eq: enhanced assumed strain} into  \cref{eq:Hu-Washizu} yields and taking all possible variations, the equilibrium constraints of the shell structure can be obtained
\begin{equation}
\begin{aligned}
\delta{\varPi}=&\int_{\Omega}\left(\frac{\partial W}{\partial\bar{\bm\varepsilon}}:\delta\bar{\bm\varepsilon}^u+\frac{\partial W}{\partial\bar{\bm\varepsilon}}:\delta\bar{\bm\varepsilon}^a\right)\text{d}{\Omega}-\mathbf{P}_e(\delta\bar{\mathbf{u}})\\
=&\int_{\Omega}\left(\bar{\bm\sigma}:\delta\bar{\bm\varepsilon}^u+\bar{\bm\sigma}:\delta\bar{\bm\varepsilon}^a\right)\text{d}{\Omega}-\mathbf{P}_e(\delta\bar{\mathbf{u}})
\end{aligned}\label{equ: ddc_shell_virtual work}
\end{equation}
By introducing the finite element discretization and distance function (shown in \cref{DD}), one can derive the formula of data-driven computing. For more details of the shell formulation, one can refer to Kuang et al. \cite{kuang2021computational}.}

\begin{figure}[htbp]
\centering
\includegraphics[width=14cm]{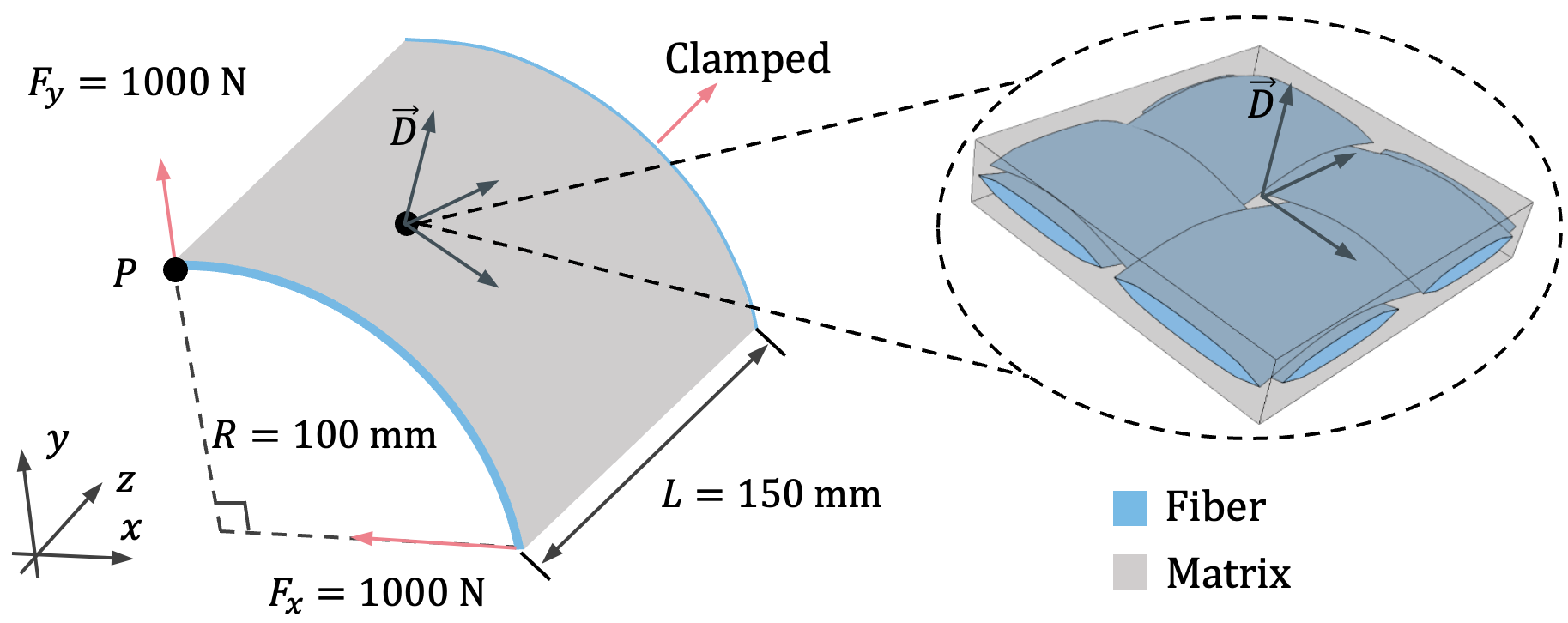}
\caption{\textcolor{black}{Sketch of the cylindrical composite shell and the corresponding RVE.}}
\label{fig: cylindrical shell}
\end{figure}

\textcolor{black}{Sketch of the shell is shown in \cref{fig: cylindrical shell}, where one side of the shell is clamped and the other side is under concentrated loads $F_x=F_y=1000~$N. The RVE of the shell is a woven fabric reinforced composite, and its geometry is obtained from microtomographic images of the composite scanned by Micro-CT \cite{huang2021data}. The parameters of the composite material are shown in \cref{table:woven fabric}. At each material point on the shell surface, the normal direction of the RVE, marked as $\vec{D}$ in \cref{fig: cylindrical shell}, is always aligned with that of the shell surface at this material point. The reference solution is obtained by the multiscale finite element method (classical FE$^2$ \cite{raju2021review}), which is implemented on the commercial software ABAQUS. The dimension of the material data for this problem $D$ is 12, and the parameters $n$ and $n_m$ for ZNE are set to $7$ and $10^{10}$, respectively.}

\begin{table}[t]
\caption{\textcolor{black}{Material parameters of the woven fabric reinforced composite.}}
\centering
\renewcommand{\arraystretch}{1.5}
\renewcommand\tabcolsep{12.0pt}
\footnotesize{
\begin{tabular}{cccccc}
\toprule
 & $E_1$ (MPa) &  $E_2=E_3$ (MPa) &  $G_1=G_2=G_3$ (MPa) & $\mu_{12}$  & $\mu_{13}=\mu_{23}$ \\
\hline 
Fiber    &   $1.18\times 10^5$   & $1.07\times 10^4$     &   $5\times 10^3$   &     $0.3$      &        $0.4$           \\ 
Matrix     &   $3\times 10^3$   & $3\times 10^3$     &   $1.25\times 10^3$   &     $0.2$      &        $0.2$          \\ 
\bottomrule
\end{tabular}}
\label{table:woven fabric}
\end{table}

\begin{figure}[t]
\centering
\includegraphics[width=\textwidth]{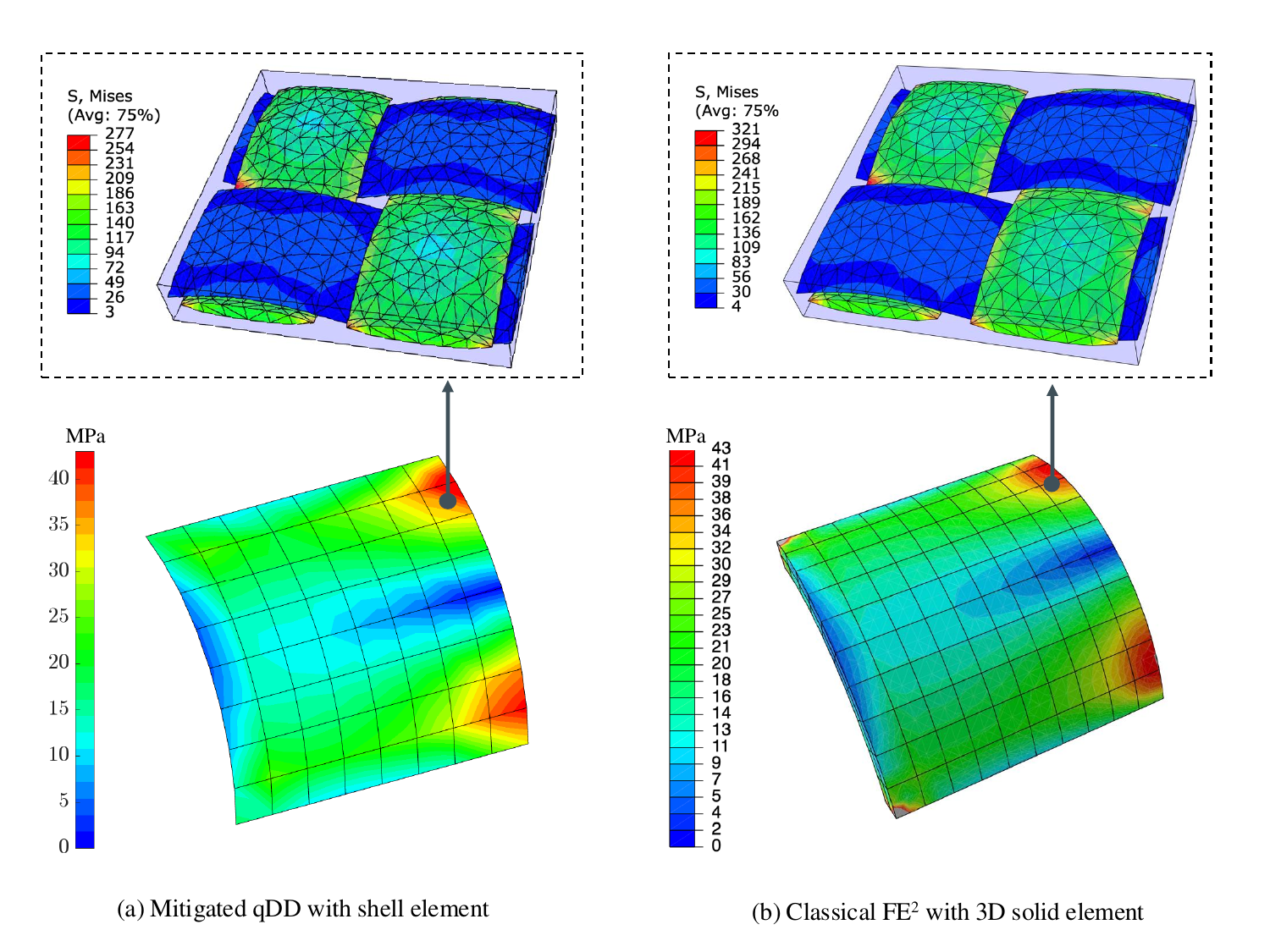}
\caption{\textcolor{black}{Von Mises stress fields of the shell surface and the RVE simulated by (a) mitigate qDD and (b) classical FE$^2$.}}
\label{fig: shell_result}
\end{figure}

\textcolor{black}{
The von Mises stress of both the shell surface and the RVE at the point of stress concentration is presented in \cref{fig: shell_result}. To facilitate comparison at the macroscopic scale, the range of stress fields obtained from both methods is set as $[0, 43]$ MPa. The results demonstrate that the mitigated qDD accurately predicts the stress distribution on the shell surface, though the predicted maximum stress of the RVE shows some discrepancy compared to the reference solution (277 MPa for the mitigated qDD and 321 MPa for the classical FE$^2$).
Furthermore, the displacement $u_y$ at the loading point $P$, indicated in \cref{fig: cylindrical shell}, is predicted as 2.30 mm for the mitigated qDD and 2.19 mm for the classical FE$^2$, yielding a relative error of approximately 5\% for the mitigated qDD. These findings further validate the effectiveness of the mitigated qDD in the 3D case. }

%
%
\section{Conclusion}\label{Conclusion}

In realizing practical applications of quantum computing in data-driven computational {homogenization}, hardware noise is an unavoidable problem for NISQ quantum computers. 
We use zero-noise extrapolation (ZNE) to deal with this issue, which helps increase the accuracy of distance calculations and makes quantum computing perform better in data-driven computational homogenization. Two quantum algorithms for distance calculation are investigated, i.e., the Swap-based and the H-based algorithms, both of which can exponentially reduce the computational complexity. We apply ZNE in these two quantum algorithms, and the results show that the accuracy of both is improved and reaches a similar level. However, the H-based algorithm requires fewer qubits and quantum gates, exhibiting an advantage for NISQ quantum computers. Furthermore, we combine this quantum algorithm with the $k$-d tree data structure, making the distance calculations in data-driven computing even more efficient. Validation for a roof structure example demonstrates the effectiveness of ZNE in improving the accuracy of data-driven computing. \textcolor{black}{Finally, applications to a 2D composite L-shaped beam and a 3D composite cylindrical shell demonstrate the effectiveness of error-mitigated quantum computing in data-driven computational homogenization.}

 
%

{\color{black}We have identified several limitations of the proposed method and outlined potential research directions that warrant further exploration in the future.} First, a real quantum computer that supports deep circuit depth is favored. After folding the gates in the circuits multiple times, the depth of the circuits may exceed the supported circuit depth of a current quantum computer. Therefore, error mitigation experiments for distance calculation on real quantum computers call for more advanced hardware devices. Second, 
we might explore new extrapolation methods \cite{van2023probabilistic,kim2023evidence}, or even different error mitigation techniques like probabilistic error cancellation (PEC) \cite{temme2017error}. {Thirdly}, this work only demonstrates error mitigation concerning state preparation with quantum gates since qRAM is not yet available \cite{aaronson2015read, biamonte2017quantum}. The error mitigation for qRAM deserves further investigation, which could be a challenging task considering the requirements of an exponential number of ancillary qubits and long coherence time \cite{giovannetti2008quantum,giovannetti2008architectures}. 
{\color{black}Fourthly}, the proposed method could be applied in the analysis of composite materials \textcolor{black}{containing multiple phases \cite{fantuzzi2020three} or randomly distributed inclusions with random shapes in RVEs \cite{kohlhaas2015fe,savvas2014homogenization,aranda2021multiscale}, aiming to
}demonstrate its performance in real-world engineering problems.
{\color{black}Finally, extending the proposed method to more complex computational tasks, including dynamics \cite{kirchdoerfer2018data,gebhardt2020framework}, inelasticity \cite{eggersmann2019model}, and multi-physics coupling problems \cite{marenic2022data}, would further leverage the full potential of quantum computing.}


%

%
%
\section*{Acknowledgements}

This work has been supported by the National Key R\&D Program of China (Grant No. 2022YFE0113100) and the National Natural Science Foundation of China (Grant No. 12432009, 11920101002, 12172262 and 12202322).

%
%
\appendix
\renewcommand\thesection{\appendixname~\Alph{section}}

{\color{black}
\section{Microscopic problem}\label{sec: micro}
{In the microscopic problem, a representative volume element (RVE) is considered, and the strain-stress data $(\bar{\bm\varepsilon},\bar{\bm\sigma})$ is collected through computational homogenization on the RVE. Specially, the macroscopic strain $\bar{\bm{\varepsilon}}$ from each material data is applied to the RVE in the form of periodic boundary conditions \cite{hui2019multiscale}, and the resulting microscopic stresses $\bm\sigma$ are averaged to obtain the equivalent macroscopic stress $\bar{\bm\sigma}$. } The microscopic Cauchy stress $\bm{\sigma}$ and its conjugate strain $\bm{\varepsilon}$ are adopted, and the microscopic homogenization problem is formulated by
\begin{subequations}\label{Eq5}
	\begin{align}
		&\int_{\omega}  {\bm{\sigma} :\delta \bm{\varepsilon} \text{d} \omega  = 0}\label{Eq5a} \\
		&\rm d\bm{\sigma}=\mathbb{C}_{t}^{(r)}:\rm d\bm{\varepsilon}\label{Eq5b}\\
		&\bm{u}^{+}-\bm{u}^{-}=\bar{\bm{\varepsilon}}\cdot(\bm{X}^{+}-\bm{X}^{-})\label{Eq5c}\\
		&\bar{\bm{\sigma}}=\frac{{1}}{{|\omega|}}\int_{\omega}{\bm{\sigma} \rm d\omega}\label{Eq5d}
	\end{align}	
\end{subequations}
where $\omega$ is the domain of the RVE, $\mathbb{C}_{t}^{(r)}$ is the constitutive tensor of each material phase $r$. 
\cref{Eq5c} is the periodic boundary condition, which loads the macroscopic strain $\bar{\bm{\varepsilon}}$ to the boundary of RVE. The displacement is denoted by $\bm{u}$, the superscripts `+' and `-' specify two opposite surfaces of the RVE, and $\bm{X}$ is the coordinate of the material point. Once $\bar{\bm{\varepsilon}}$ is given, \cref{Eq5a,Eq5b,Eq5c} can be solved using the Newton-Raphson method to get the deformed RVE. In post-processing, \cref{Eq5d} is used to compute the macroscopic Cauchy stress $\bar{\bm{\sigma}}$. In this manner, homogenized strain-stress data ($\bar{\bm{\varepsilon}},\bar{\bm{\sigma}}$) can be generated and stored in the database $\bar{\mathscr{D}}$.

{\color{black}
\section{Swap-based algorithm}\label{sec: swap}

\cref{fig: sketch_circuit} (a) presents the quantum circuit of the Swap-based algorithm for distance calculation \cite{xu2023quantum}. First, two quantum states containing the information of $\bm{V}$ and $\bm{V}'$ are encoded into the quantum computer via qRAM in time $O(\log D)$ \cite{giovannetti2008quantum,giovannetti2008architectures}
\begin{subequations}\label{states2}
\begin{align}
& |\phi\rangle=\frac{1}{\sqrt{Z}}(| \bm{V} ||0\rangle-| \bm{V}' ||1\rangle)\\
& |\psi\rangle=\frac{1}{\sqrt{2}}(|0\rangle|\bm{V}\rangle+|1\rangle|\bm{V}'\rangle)
\end{align}
\end{subequations}
where the Dirac notation $|\cdot \rangle$ denotes a quantum state, and $Z=| \bm{V} |^2+| \bm{V}'|^2$. In addition, the amplitudes of the quantum states $|\bm{V}\rangle$ and $|\bm{V}'\rangle$ represent the normalized $\bm{V}$ and $\bm{V}'$, respectively.
Second, the swap test \cite{buhrman2001quantum} is used to compute the inner product $|\langle \phi| \psi \rangle|^2$, which includes two Hadamard gates and one $\textit{CSWAP}$ gate. Note that the swap test consists of only three gates and has a time complexity of $O(1)$, independent of $D$. At the end of the circuit, the measurement result of the first qubit can be $|0\rangle$ or $|1\rangle$ due to the superposition of the quantum state, and the probability of being $|0\rangle$ is
\begin{equation}\label{p}
\begin{split}
p_s = \frac{1}{2} + \frac{1}{2}{| {\langle \phi  | \psi  \rangle } |^2}
\end{split}
\end{equation}
By substituting \cref{states2} into \cref{p}, the distance is obtained by
\begin{equation}\label{dis_final}
d=4Z(p_s-\frac{1}{2})
\end{equation}
To compute $d$ using \cref{dis_final}, we need to obtain $Z$ and $p_s$, where $Z$ is pre-computed in the classical computer \cite{xu2023quantum} and $p_s$ is calculated by running the swap test circuit. 
Due to the inherent properties of quantum systems, measuring a quantum state leads to its collapse \cite{Nielsen}, resulting in the acquisition of a single bit of classical information.
This intrinsic property necessitates the execution of the quantum circuit multiple times and the use of statistical estimate to read out ${p_s}$ \cite{xu2023quantum}. 
Suppose we have $n_m$ measurements, and the number of $|0\rangle$ in the measurement results is $n_0$. According to the maximum-likelihood estimation, $p_s$ is estimated by $\hat{p}_s=n_0/n_m$. By substituting $\hat{p}_s$ in \cref{dis_final}, we have
\begin{equation}\label{dis_final2}
\hat{d}=4Z(\frac{n_0}{n_m}-\frac{1}{2})
\end{equation}
where the estimated distance $\hat{d}$ for $d$ is obtained.

}

\section{Complexity analysis}\label{sec: complexity analysis}

Given that both the Swap-based (see \ref{sec: swap}) and H-based algorithms statistically estimate their respective probabilities, $\hat{p}_s$ and $\hat{p}_h$, there inherently exists an estimation error that impacts the accuracy of the estimated distance $\hat{d}$. This section presents an analysis of the error and time complexity for these two quantum algorithms. Specifically, we derive the relationship between the root-mean-square error (RMSE) of the estimated distance $\hat{d}$ and the number of measurements $n_m$ used in each algorithm. It is crucial to note that this error analysis is focused solely on the statistical estimation aspects and does not account for potential inaccuracies arising from quantum hardware noise.

For the Swap-based algorithm, the derivation of the squared RMSE, denoted by $\epsilon_{s}^2$, is provided in \cite{xu2023quantum}, where interested readers can find further details. The squared RMSE is given by
\begin{equation}\label{dis_final2_V}
\epsilon_{s}^2 = 16\left(| \bm{V} |^2+| \bm{V}'|^2\right)^2 \frac{p_s (1-p_s)}{n_m}
\end{equation}
Similarly, for the H-based algorithm, we derive the squared RMSE of the estimated distance $\hat{d}$ from \cref{H-d2}, denoted by $\epsilon_{h}^2$
\begin{equation}\label{H_RMSE}
{\epsilon_h}^2 =  16{| \bm{V} |}^2{| \bm{V}'|}^2 \frac{p_h(1-p_h)} {n_m} 
\end{equation}
Therefore, we have $\epsilon_s^2 \sim O\left({1}/{n_m}\right)$ and $\epsilon_h^2 \sim O\left({1}/{n_m}\right)$. Equivalently, the upper bounds of $n_m$ for the two algorithms scale the same, i.e., $n_m \sim O\left({1}/{\epsilon_s^2}\right)$ for the Swap-based algorithm and $n_m \sim O\left({1}/{\epsilon_h^2}\right)$ for the H-based algorithm. Given that the time complexities for executing the circuits once for both algorithms are $O(\log D)$, the upper bounds of the total time for these quantum algorithms are also equivalent, namely $O\left({\log D}/{\epsilon_s^2}\right)$ for the Swap-based algorithm and $O\left({\log D}/{\epsilon_h^2}\right)$ for the H-based algorithm. Hence, both quantum algorithms can exponentially reduce the computational complexity of distance calculation from $O(D)$ to $O(\log D)$.

Though the upper bounds of ${\epsilon_s}^2$ and ${\epsilon_h}^2$ scale the same, we have an interesting finding that there exists a relation ${\epsilon_h}^2 \leq {\epsilon_s}^2$, i.e., the H-based algorithm has a lower estimation error than the Swap-based algorithm under the same $n_m$ and regardless of the quantum hardware noise. Here we present the proof of this relation. From \cref{dis_final2_V,H_RMSE}, the relation ${\epsilon_h}^2 \leq {\epsilon_s}^2$ is equivalent to
\begin{equation}\label{equiv1}
16{| \bm{V} |}^2{| \bm{V}'|}^2 \frac{p_h(1-p_h)} {n_m}  \leq 16\left(| \bm{V} |^2+| \bm{V}'|^2\right)^2 \frac{p_s (1-p_s)}{n_m}
\end{equation}
By eliminating $n_m$ and replacing $p_h$ and $p_s$ using the relations in \cref{H-d,dis_final}, we have 
\begin{equation}\label{equiv2}
4|\bm{V}|^2|\bm{V}^\prime|^2 - Z^2 + 2dZ - d^2 \leq 4Z^2-d^2
\end{equation}
which can be further simplified as
\begin{equation}\label{equiv3}
4|\bm{V}|^2|\bm{V}^\prime|^2 - 4 \cos(\theta) |\bm{V}||\bm{V}^\prime|  (|\bm{V}|^2+|\bm{V}^\prime|^2)-  3(|\bm{V}|^2+|\bm{V}^\prime|^2)^2 \leq 0
\end{equation}
where $\cos(\theta)$ is the cosine of the angle between $\bm{V}$ and $\bm{V}^\prime$, and $\cos(\theta)\in[-1,1]$. Thus, the  sufficient condition of \cref{equiv3} is
\begin{equation}\label{equiv4}
4|\bm{V}|^2|\bm{V}^\prime|^2 + 4 |\bm{V}||\bm{V}^\prime| (|\bm{V}|^2+|\bm{V}^\prime|^2)-  3(|\bm{V}|^2+|\bm{V}^\prime|^2)^2 \leq 0
\end{equation}
Without loss of generality, we set $|\bm{V}^\prime|=k|\bm{V}|$, where $k\geq0$. Then \cref{equiv4} can be simplified as
\begin{equation}\label{equiv5}
-3k^4+4k^3-2k^2+4k-3 \leq 0
\end{equation}
which can be easily proved since the left-hand side only has one critical point at $k=1$ and the corresponding value is $0$. Therefore, we have proved that ${\epsilon_h}^2 \leq {\epsilon_s}^2$. We would like to mention that this proof only demonstrates that the H-based algorithm has a lower estimation error without the consideration of quantum hardware noise. When the quantum hardware noise is considered, this advantage may not be obvious, as illustrated in \cref{sec:Validation of distance}.

}

\section{Quantum noise model}\label{append:noise_model}
In this paper, a noise model \cite{Qiskit,blank2020quantum} is employed to approximate the hardware noise of real devices, which includes two parts:
\begin{itemize}
    \item Single-qubit gate errors consisting of a single-qubit depolarizing error and a thermal relaxation error.
    \item Two-qubit gate errors consisting of a two-qubit depolarizing error followed by single qubit thermal relaxation error on each qubit participating in the gate. 
\end{itemize}

Thermal relaxation is a non-unital (i.e., irreversible) process that describes the thermalization of the qubit spins towards a ground state $\left|0\right\rangle$ or an excited state $\left|1\right\rangle$. This noise model involves two different expressions which depends on the regime of relaxation time $T_1$ and dephasing time $T_2$. For the case $T_2 \leq T_1$,
the thermal relaxation model is implemented as a mixture of 
$I$, $Z$, reset to $\left|0\right\rangle$, reset to $\left|1\right\rangle$

\allowdisplaybreaks
\begin{align}
&\rho \xrightarrow{~~~~} q_{\mathrm{id}}\rho + q_z Z\rho Z+ q_{r_0}\left|0\right\rangle\left\langle0\right|\rho \left|0\right\rangle\left\langle0\right|+q_{r_1}\left|1\right\rangle\left\langle1\right|\rho \left|1\right\rangle\left\langle1\right|\nonumber\\
&q_{\mathrm{id}}=1-q_{z}-q_{r_0}-q_{r_1}, \nonumber\\
&q_{z}=\epsilon_{T_{1}}\left(1-\epsilon_{T_{2}}\epsilon_{T_{1}}^{-1}\right)/2, \label{eq: thermal_relax}\\
&q_{r_{0}}=(1-q_{e})(1-\epsilon_{T_{1}}), \nonumber\\
&q_{r_1}=q_e(1-\epsilon_{T_{1}})\nonumber
\end{align}
where $\epsilon_{T_{1}}=e^{-T_g/T_1}$ and $\epsilon_{T_{2}}=e^{-T_g/T_2}$ are respectively the probabilities for each qubit to relax and dephase after a quantum gate with evolution time $T_g$, $q_e$ is related to the quantum processor's temperature $\Theta$, Planck's constant $h$, Boltzmann's constant $k_B$ and frequency of the qubit $f$
\begin{equation}
q_e=\left(1+e^\frac{2hf}{k_B\Theta}\right)^{-1}
\end{equation}
Since $\Theta$ is about $15~\mathrm{mK}$ , $q_e$ is close to 0 and the error of reset to $\left|1\right\rangle$ can be omitted. Therefore, \cref{eq: thermal_relax} can be simplified to
\begin{equation}
\begin{aligned}
&\rho \xrightarrow{~~~~}  q_{\mathrm{id}}\rho + q_z Z\rho Z+ q_{r_0}\left|0\right\rangle\left\langle0\right|\rho \left|0\right\rangle\left\langle0\right|\\
& q_{\mathrm{id}} = 1-q_z-q_{r_0}\\
& q_z=\left(1-q_{r_0}\right)\left(1-\epsilon_{T_2} \epsilon_{T_1}^{-1}\right) / 2 \\
& q_{r_0} =1-\epsilon_{T_1}
\end{aligned}
\end{equation}
When $2T_1>T_2>T_1$, the thermal relaxation noise model is described by a Choi-matrix representation \cite{blank2020quantum}
\begin{equation}
\rho \xrightarrow{~~~~}  \mathrm{tr}_1[\Lambda(\rho^T\otimes I)]
\end{equation}
where $\mathrm{tr}_1$ is the trace over the main system in which the density matrix $\rho$ resides and $\Lambda$ is expressed as
\begin{equation}
\Lambda=\begin{pmatrix}1-p_ep_r&0&0&\epsilon_{T_2}\\0&p_ep_r&0&0\\0&0&(1-p_e)p_r&0\\\epsilon_{T_2}&0&0&1-(1-p_e)p_r
\end{pmatrix}\xrightarrow{p_e\approx 0}\begin{pmatrix}1&0&0&\epsilon_{T_2}\\0&0&0&0\\0&0&p_r&0\\\epsilon_{T_2}&0&0&1-p_r\end{pmatrix}
\end{equation}

The depolarizing model includes a single-qubit depolarizing model and a two-qubit depolarizing model. The former is expressed as
\begin{equation}
\rho\xrightarrow{~~~~} (1-q_1)\rho+\frac {q_1}{4}\sum_{i=0}^{3}E_i\rho E_i^{\dagger}
\end{equation}
while the latter is represented as
\begin{equation}
\rho\xrightarrow{~~~~} (1-q_2)\rho+\frac{q_2}{16}\sum_{i=0}^{3}\sum_{j=0}^{3}E_{ij}\rho E_{ij}^{\dagger}, E_{ij} = E_{i}\otimes E_{j}
\end{equation}
where $E$ contains the identity gate $I$ and Pauli gates $\{X, Y, Z\}$
\begin{equation}
E_{0} =I, E_{1} = X, E_{2} = Y, E_{3} = Z  
\end{equation}
The depolarizing probabilities $q_1$ and $q_2$ are correlated with the gate error rate $\epsilon_g$, the relaxation error rate $\epsilon_{T_1}$ and dephasing error rate $\epsilon_{T_2}$. Assuming that all the qubits have the same relaxation and dephasing time, then $q_1$ and $q_2$ can be expressed as
\begin{equation}
\begin{aligned}
&q_1=1+3\frac{2\epsilon_g-1}{d_1}, d_1 = \epsilon_{T_1}+2\epsilon_{T_2}\\
&q_2=1+5\frac{4\epsilon_g-3}{d_2}, d_2 = 2\epsilon_{T_1}+\epsilon_{T_1}^2+4\epsilon_{T_2}+4\epsilon_{T_2}^2+4\epsilon_{T_1}\epsilon_{T_2}
\end{aligned}
\end{equation}

The required device data are calibrated from the IBM quantum computer \textit{ibm\_osaka} on 2024-04-15 12:15 UTC, and median values are used for noise parameters, i.e., $T_1 = 280~\rm{\mu s}$, $T_2 = 127~\rm{\mu s}$, gate time $T_g = 0.06~\rm{\mu s}$ and error rate $\epsilon_g = 2.77\times 10^{-4} $ for 1-qubit gates $I$, $X$ and $SX$, gate time $T_g = 0.66~\rm{\mu s}$ and error rate $\epsilon_g = 8.56\times 10^{-3}$ for a two-qubit gate $ECR$.
%

\section{Normal approximation}\label{sec: Normal}

To estimate the probability $\hat{p}=n_0/n_m$, we need to run the quantum circuit $n_m$ times, and then get the number of $|0\rangle$ in the measurement results $n_0$. This process is time-consuming when numerically conducted on a quantum computer simulator. 
Since one measurement relates to state $|0\rangle$ with probability $p$ and state $|1\rangle$ with probability $1-p$, $n_0$ follows a binomial distribution $n_0 \sim B(n_m, p)$. 
According to the central limit theorem, the distribution of $n_0$ becomes approximately normal with $n_0 \sim N(n_mp, n_mp(1-p))$ if both $n_mp$ and $n_m(1-p)$ are larger than 5. Hence, when the estimation of $p$ is performed through a quantum computer simulator, it can be achieved by generating a random variable that obeys the corresponding normal distribution, which reduces the computational complexity of obtaining $\hat p$ from $O(n_m)$ to $O(1)$.

\begin{figure}[tbp]
\centering
\subfloat[Distance distribution]{
\centering
\label{fig: surrogate_counts}
\includegraphics[width=0.45\linewidth]{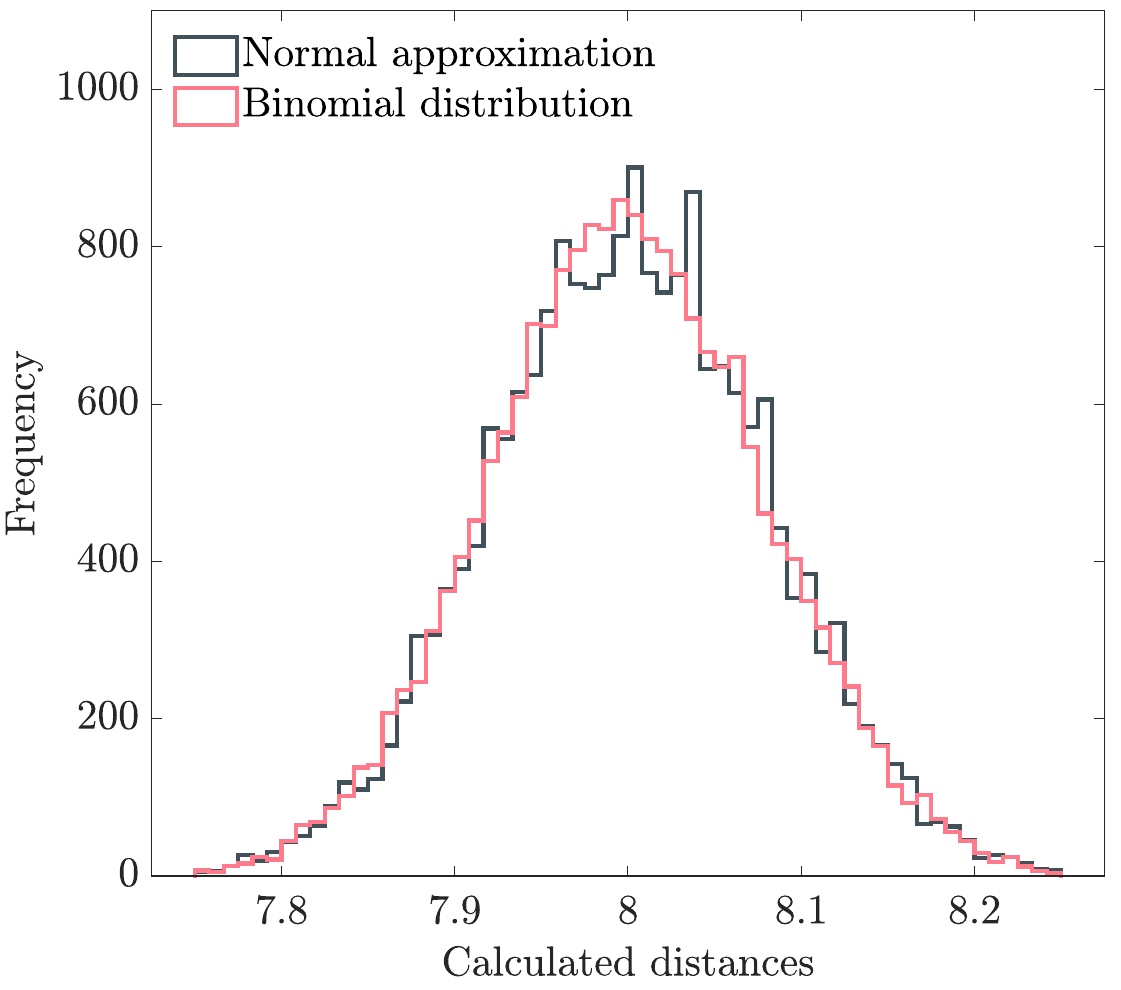}}
\subfloat[CPU time]{
\centering
\label{fig: surrogate_times}
\includegraphics[width=0.45\linewidth]{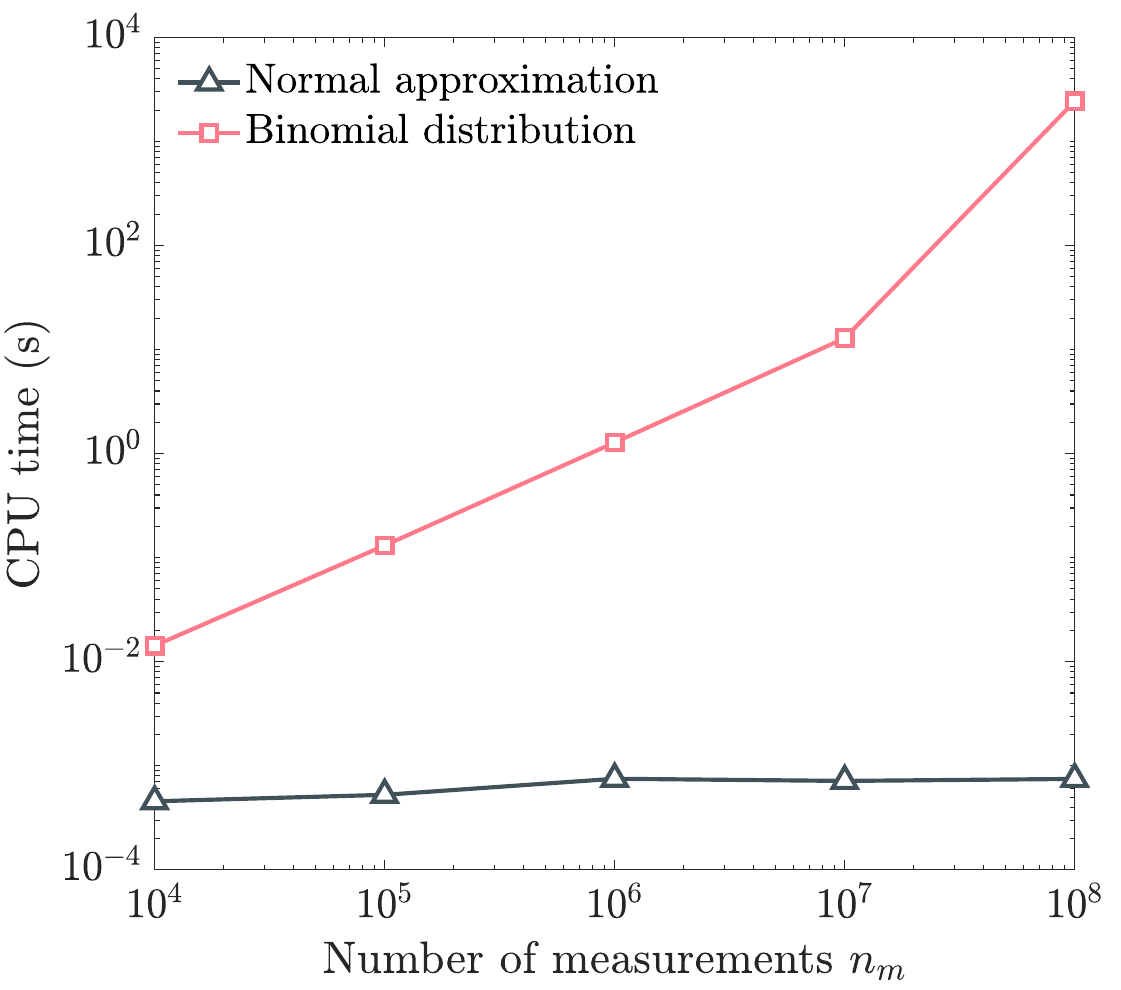}}
\caption{Performance of the normal approximation in comparison to the binomial distribution.}
\label{fig: surrogate model}
\end{figure}

Two numerical tests are conducted to verify the effectiveness and efficiency of the estimation method with normal approximation. Firstly, the distances between vectors $(0,2)$ and $(2,0)$ are calculated based on normal approximation and binomial distribution, where the number of measurements is set to $n_m=10^5$ and $20000$ samples of distances are included. As shown in \cref{fig: surrogate model} (a), the distance distribution obtained by normal approximation is similar to the one obtained by the binomial distribution. This means the estimation method with normal approximation retains the same statistical characteristics as the one with binomial distribution. Then, the computational times of the two estimation methods versus the number of measurements $n_m$ are tested. As shown in \cref{fig: surrogate model} (b), the required computational time for normal approximation is independent of $n_m$, which shows its significant advantage in efficiency.

\bibliographystyle{elsarticle-num}
\bibliography{qDD_COM}
\end{document}